\documentclass[pre, twocolumn,aps,showpacs]{revtex4}
\usepackage{graphicx}
\usepackage{dcolumn}
\usepackage{bm}
\usepackage{color}
\usepackage{amsmath}
\usepackage{amssymb}
\usepackage{comment}

\begin{document}

\title{Conformations, Transverse Fluctuations and Crossover Dynamics of a Semi-Flexible Chain in Two Dimensions}
\author{Aiqun Huang}
\author{Aniket Bhattacharya}
\affiliation{Department of Physics, University of Central Florida, Orlando, Florida 32816-2385, USA}
\altaffiliation[]{Author to whom the correspondence should be addressed}
\email{aniket@physics.ucf.edu}
\author{Kurt Binder}
\affiliation{Institut f\"ur Physik, Johannes Gutenberg-Universit\"at Mainz, 
Staudinger Weg 7, 55099 Mainz, Germany}
\date{\today}

\begin{abstract}
We present a unified scaling description for the dynamics of monomers of a semiflexible chain under good solvent condition 
in the free draining limit. We consider both the cases where the contour length $L$ is comparable to the persistence 
length $\ell_p$ and the case $L\gg \ell_p$. Our theory captures the early time monomer dynamics of a stiff chain characterized by $t^{3/4}$
dependence for the mean square displacement(MSD) of the monomers, but predicts a first crossover to the Rouse regime of $t^{2\nu/{1+2\nu}}$ for 
$\tau_1 \sim \ell_p^3$, and a second crossover to the purely diffusive dynamics for the entire chain at $\tau_2 \sim L^{5/2}$.
We confirm the predictions of this scaling description by studying monomer dynamics of dilute solution of 
semi-flexible chains under good solvent conditions obtained from our Brownian dynamics (BD) simulation studies for a large choice 
of chain lengths with number of monomers per chain N = 16 - 2048 and persistence length 
$\ell_p = 1 - 500$ Lennard-Jones (LJ) units. These BD simulation results further confirm the absence of Gaussian regime for a 2d swollen 
chain from the slope of the plot of $\langle R_N^2 \rangle/2L \ell_p \sim  L/\ell_p$ which around $L/\ell_p \sim 1$ changes suddenly from 
$\left(L/\ell_p \right)  \rightarrow \left(L/\ell_p \right)^{0.5} $, also manifested in the power law decay for the bond autocorrelation 
function disproving the validity of the WLC in 2d.  We further observe that the normalized transverse fluctuations of the 
semiflexible chains for different stiffness $\sqrt{\langle l_{\bot}^2\rangle}/L$ as a function of renormalized 
contour length $L/\ell_p$ collapse on the same master plot and exhibits power law scaling 
$\sqrt{\langle l_{\bot}^2\rangle}/L \sim (L/\ell_p)^\eta $ at extreme limits, where $\eta = 0.5$ for extremely stiff chains 
($L/\ell_p \gg 1$), and $\eta = -0.25$ for fully 
flexible chains. Finally, we compare the radial 
distribution functions obtained from our simulation studies with those obtained analytically.    
\end{abstract} 
\maketitle

\section{Introduction}
Macromolecules adsorbed on substrate surfaces occur in many different contexts, from materials science to biophysics, 
{\em e.g.}, biomolecules interacting with cell membranes. Hence the understanding of conformations and dynamics of 
macromolecules in such a (quasi-) two-dimensional geometry has been of long-standing interest \cite{deGennes_1987,Fleer,Eisenriegler,Netz_2003}. 
Note also that many methods to characterize polymer conformations experimentally, {\em e.g.}, electron 
microscopy \cite{Takabayashi,Stokke,Trachtenberg,Bednar,Agaskova,Papadopoulos,Lehman}, atomic force microscopy (AFM) \cite{Hansma,Yoshinaga,Valle,Moukhtar_PRL_2005,Rechendorff,Langowski,Moukhtar_2010}, 
fluorescence microscopy of suitably labelled biopolymers \cite{Maier_PRL_1999,Maier_2000}, require that these macromolecules 
are attached to a substrate. In this context, considerations of macromolecules confined to a strictly 
two-dimensional geometry are of interest, at least as a limiting case. The same statement holds 
when one considers macromolecules confined to nanoslits with non-adsorbing walls, a topic that 
also has found much recent interest.

While the statistical mechanics of completely flexible polymers in $d=2$ dimensions has been studied 
extensively since a long time and is well understood \cite{deGennes_book,Sokal}, under many circumstances it should 
be taken into account that macromolecules are stiff and not flexible on small scales \cite{Flory,Grosberg,Rubinstein}. 
This is true both for simple synthetic polymers {\em e.g.}, polystyrene, alkane  chains, {\em etc.}, and for 
various biopolymers, {\em e.g.}, double-stranded (ds) and single stranded (ss) DNA, polysaccharides, proteins, {\em etc.} \cite{Parry}. 
Apart from very stiff polymers ({\em e.g.}, Actin, Titin, microtubules, {\em etc.}) the ``persistence 
length $\ell_p$ \cite{Flory,Grosberg,Rubinstein,Parry,Hsu_Macro_2010} characterizing the stiffness typically is much less than the contour 
length $L$ of a macromolecule, and the crossover from rod-like behavior to the behavior of flexible 
polymers needs to be considered. As is well-known, the scales of interest range
from the sub-nanometer scale to the micrometer scale \cite{Binder_book,Thirumalai_Nature_2011}, and hence in the theoretical 
modeling coarse-grained models must be used \cite{Binder_book,Thirumalai_Nature_2011,Hsu_Macro_Simul_2011}.\par

In the present work, we wish to address the problem of polymer conformation and dynamics for semi-flexible polymers 
in two dimensions, using Molecular Dynamics simulations of a bead-spring type model with a bond angle potential 
by which we can control the stiffness of the chains over a wide range. We note that the standard analytical coarse-grained 
description in terms of the Kratky-Porod \cite{Kratky,Harris} model for wormlike chains in $d=2$ dimensions is not very 
useful, since it neglects excluded volume effects completely, although they are known to be very important in $d=2$ \cite{Hsu_EPL_2011}. 
A study of semiflexible polymers in terms of a lattice model \cite{Hsu_EPL_2011,Hsu_JCP_2012_1,Hsu_JCP_2012_2,Hsu_Soft_2013} 
is expected to yield valid results for universal properties of semiflexible polymer, {\em i.e.}, on length scales much 
larger than the persistence length; but on smaller scales it can describe only stiffness of the type similar to that of 
alkane chains, where $\ell_p$ is of the order of the typical length of ``all trans'' sequences, in between monomers taking 
a gauche ($g\pm$) minimum in the torsional potential. For such cases {\em e.g.}, dsDNA we expect that the (small) 
flexibility of the macromolecules is due to fluctuations in bond  lengths and bond angles, rather than disorder 
in the population of states in the torsional potential, and then the present off-lattice model is more realistic. In addition, 
the lattice work \cite{Hsu_EPL_2011,Hsu_JCP_2012_1,Hsu_JCP_2012_2,Hsu_Soft_2013} applying the pruned enriched Rosenbluth method 
(PERM)~\cite{Grassberger_1997,Hsu_Stat_2011} could not address the dynamics of the chains at all. Previous work on the dynamics 
of single semiflexible chains in dilute 
solution~\cite{Pincus_MM_1980,Harnau2,Kroy_PRE_1997,Shusterman_PRL_2004,Petrov,Netz_EPL_2009,Hinczewski,Hinczewski_PhysicaA,Steinhauser} 
has focused on the case $d=3$ almost exclusively, and most of the 
work~\cite{Pincus_MM_1980,Harnau2,Kroy_PRE_1997,Shusterman_PRL_2004,Petrov,Netz_EPL_2009,Hinczewski,Hinczewski_PhysicaA} 
has studied the effect of hydrodynamic (HD) interactions mediated by the solvent. 
Assuming that the substrate surface provides a stick boundary condition with respect to solvent fluid 
flow, one can show \cite{Winkler} that HD interactions are essentially screened, and hence 
are ignored here (as well as in our preliminary communication where a small part of our results were 
presented \cite{Huang_epl}) from the outset.\par

While the Kratky-Porod Worm-like-chain (WLC) Model has been found to be grossly inadequate to describe a semiflexible chain 
in 2d, vast amounts of analytical and numerical work have been accumulated using the WLC model as the 
starting point~\cite{Harnau1,Harnau2,Winkler_JCP_2003}.  
Recent experimental results of confined biopolymers on a 2d substrate are also analyzed using 
the well known results of WLC model~\cite{Moukhtar_PRL_2005}. Therefore,
in the following section we summarize the main results of the WLC model which we will revisit in the subsequent sections
to compare our simulation results. The organization of the paper is as follows. In the next section 
we introduce the WLC model. Next in Sec.~\ref{scaling_section} a scaling theory is derived where we show that monomer dynamics 
of a semiflexible polymer exhibits a double crossover as a function of time. 
We then introduce the bead spring model for a semiflexible chain in Sec.~\ref{model_section}.
The results of BD simulation are presented in Sec.~\ref{BD_section}, which is divided into two sub-sections:
The equilibrium properties are presented in Sec.~\ref{eqlb_section}; in Sec.~\ref{dynamics_section} among other 
results we validate the predictions
of the scaling theory using (BD) simulation for chains of different length and stiffness. 
\section{Kratky-Porod Worm-like-chain (WLC) Model}
\label{WLC_section}
The Hamiltonian corresponding to the bending 
energy for the WLC model is given by
\begin{equation}       
\mathcal H = \frac{\kappa}{2} \int_0^L \left (\frac{\partial^2 \mathbf{r}}{\partial s^2}\right)^2ds,
\label{wlc}
\end{equation}
where $\mathbf{r}(s)$ is the position vector of a mass point, $L$ is the inextensible contour length, $\kappa$ is the bending rigidity, 
and the integration is carried out along the contour $s$~\cite{Rubinstein,Doi}. 
Using symmetry arguments for the free energy it can be shown~\cite{Landau} that the chain persistence 
length $\ell_p$ for a WLC in 2d and 3d are given by\\

\begin{subequations}
\begin{minipage}{0.45\textwidth}
\begin{align}
\label{lp2d}
\ell_ p &=\frac{2\kappa}{k_BT}\;\; {\rm (2d);~~}
\end{align}
\end{minipage}
\begin{minipage}{0.45\textwidth}
\begin{align}
\label{lp3d}
\ell_p &=\frac{\kappa}{k_BT} \;\;{\rm (3d)~~}
\end{align}
\end{minipage}
\end{subequations}\\
\\
The model has been studied quite extensively applying path integral and other 
techniques~\cite{Harnau1,Winkler_JCP_2003,Yamakawa,Wilhelm_PRL_1996} and exact expressions 
of various moments of the distribution of monomer distances along the chain have been worked out.
The end-to-end distance in the WLC model is given by~\cite{Rubinstein}
\begin{equation}
\frac{\langle R_N^2\rangle}{L^2} = \frac{2\ell_p}{L}\left(1-\frac{\ell_p}{L}[1-\exp(-L/\ell_p)]\right).
\label{rn_wlc}
\end{equation}
In the limit $\ell_p \ll L$ one gets $\langle R_N^2\rangle = 2\ell_pL$ and the chain behaves like a Gaussian coil;  
for  $\ell_P \gg L$, $\langle R_N^2\rangle = L^2$ and the chain behaves like a rod.
Evidently the model neglects the excluded volume (EV) (Eqn.~\ref{LJ} of the bead-spring model in Sec.-\ref{model_section}) interaction and hence 
interpolates between rod and Gaussian limits only. \par
The WLC model can be viewed as a limiting case of a freely rotating chain~\cite{Rubinstein}, where the correlation 
between bond vectors $\vec{b_i}$ and $\vec{b}_{i+s}$ is assumed to follow 
\begin{equation}
\left< \vec{b}_i \cdot \vec{b}_{i+s}\right> = b^2 \exp(-s/\ell_p), 
\label{bexp}
\end{equation}
where $|\vec{b_i}|=|\vec{b}_{i+s}|=b$ and the characteristic length is defined as the persistence length $\ell_p$. From Eqn.~\ref{bexp} it then immediately 
follows that $\ell_p$ can be calculated from the bond angle $\cos \theta = \hat{b}_i \cdot \hat{b}_{i+1} $, where $\hat{b}_i$ is the unit vector of 
the corresponding bond vector $\vec{b}_i$  as follows:
\begin{equation}
\ell_p = -\frac{1}{\ln \left( \cos \theta \right)}.
\label{lp1}
\end{equation}
\par

Likewise, dynamics of the WLC model have been explored using Langevin type of 
equation~\cite{Granek_1997,Maggs_MM_1993,Kroy_PRE_1997,Wilhelm_PRL_1996,Bullerjahn_EPL_2011}. 
One can expect that the dynamics of a stiff chain will be dominated by transverse fluctuations (bending modes)~\cite{Winkler_JCP_2003} and that  
the short time dynamics will be governed by the chain persistence length. Indeed a relaxation dynamics using the WLC 
Hamiltonian (Eqn.~\ref{wlc}) approach yields an expression for fluctuation 
\begin{equation}
\langle \left( \Delta h\right)^2  \rangle \sim \ell_p^{-0.25}t^{0.75}.
\label{0.75}
\end{equation}
which crosses over to simple diffusion at late time~\cite{Granek_1997,Maggs_MM_1993}.
As we will see later from our results that even for a Gaussian chain a more ``complete theory'' should have captured 
an intermediate regime characterized by a growth law $t^{0.5}$ for a fully flexible chain for an intermediate time 
when the fluctuation becomes of the order of radius of gyration of the chain. However, the Langevin theories for the 
WLC chain did not describe this regime. 
\section{Scaling description}
\label{scaling_section}
\begin{figure}[ht!]                
\centering
\includegraphics[width=0.87\columnwidth]{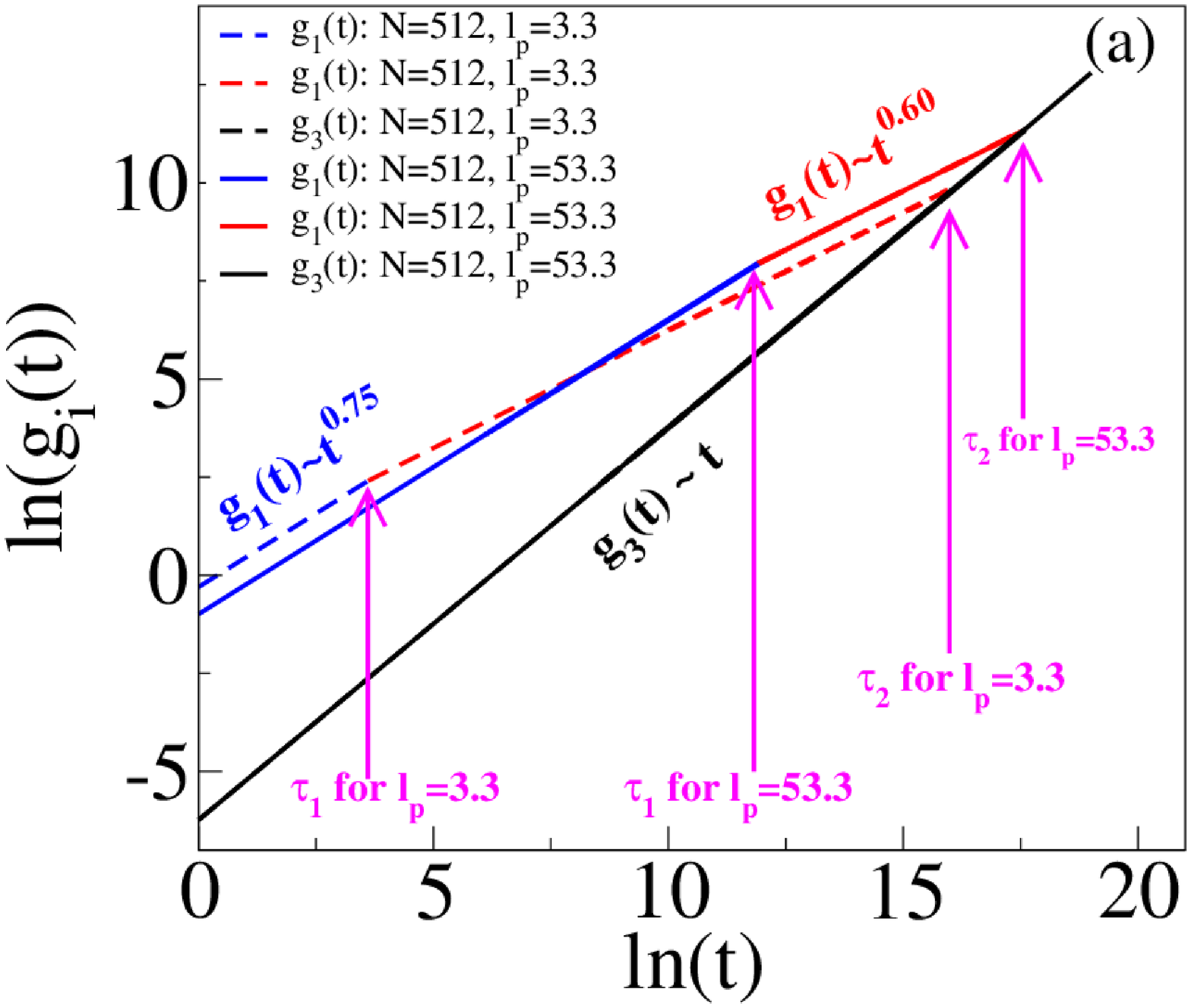}\\
\includegraphics[width=0.87\columnwidth]{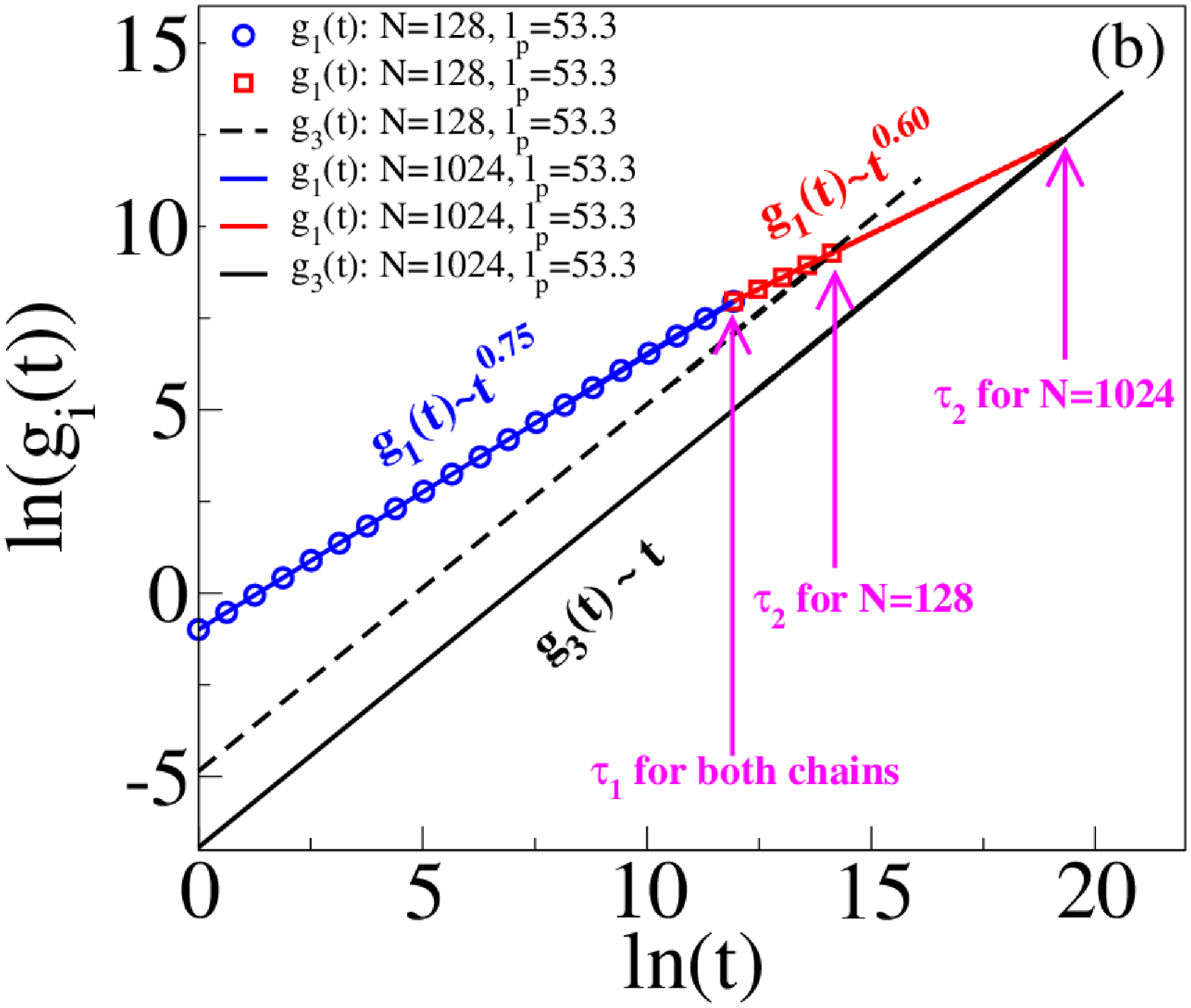}
\caption{\small Theoretical scaling for $(N,\kappa)\equiv (512,2)$, $(512,32)$(a) and 
$(N,\kappa)\equiv (128,32)$, $(1024,32)$(b). Blue (red) symbols, solid and dashed lines correspond to $g_1(t) \sim t^{0.75}$($g_1 \sim t^{0.60}$), 
black dashed, solid lines correspond to $g_3(t) \sim t$. Here the power laws Eqn.~\ref{Granek}, ~\ref{gmid} and \ref{g3_scaling} are 
plotted, using units of $b=1$, $l_p=2\kappa/k_BT$, $W=1$. The width of each region shows how these regimes depend on $\ell_p$ and $N$. 
Note that in reality we expect a very gradual change of slope on the log-log plot at both crossover times, rather than sharp kinks.}
\label{phase}
\end{figure}

We first develop the scaling description for the dynamics of a two dimensional semiflexible chain in the free draining limit.  
The free draining limit is of particular interest in 2d because it often satisfies the experimental conditions, such 
as DNA confined in a 2d substrate where the effect of hydrodynamics is negligible. For a WLC several theories 
based on Langevin dynamics have been developed most of which indicate a $t^{0.75}$ dependence of the transverse fluctuation 
of the MSD $g_1(t)$ (see Eqn.~\ref{g1}) with time $t$ for the stiff chain. Therefore,  
we start with the Eqn.~\ref{Granek} below derived by Granek~\cite{Granek_1997} and Farge and Maggs~\cite{Maggs_MM_1993} using a Langevin dynamics framework for  
the WLC Hamiltonian
\begin{equation}
g_1(t) = b^2\left(b/\ell_p\right)^{1/4}\left(Wt\right)^{3/4},
\label{Granek}
\end{equation} 
where $W$ is the monomer reorientation rate. Prefactors of order unity are omitted throughout~\cite{Comment_g1}.  
For early time the monomer dynamics will be independent of the chain length $N$
until the fluctuations in monomer position become of the order of $\ell_p$. Therefore, denoting the time when the first 
crossover occurs as $\tau_1$ and substituting 
$g_1 = \ell_p^2$ and $t= \tau_1$ in Eqn.~\ref{Granek} we immediately get 
\begin{equation}
W\tau_1 = \left(\ell_p/b\right)^3.
\label{tau1}
\end{equation}
For $ 0 < t \le W^{-1}(\ell_p/b)^3$ the monomer dynamics is described by $g_1(t) \sim t^{0.75}$ according to Eqn.~\ref{Granek} until $g_1(t) = \ell_p^2$ at time $W^{-1}(\ell_p/b)^3$.  
The width of this region is independent of $N$ and solely a function of $\ell_p$~(see Fig.~\ref{phase}). \par 
For $ \tau_1 < t < \tau_2$ the dynamics is governed by the Rouse relaxation of monomers of a fully flexible EV chain in 2d characterized by 
$g_1(t) \propto t^{2\nu/(1+2\nu)} = t^{0.6}$. $\tau_2$ characterizes the onset of  
the purely diffusive regime when $g_1(\tau_2) = \langle R_N^2 \rangle$~\cite{Grest_Kremer_PRA_1986}. Recall that the exponent $\nu$ 
that describes the scaling of the end-to-end distance $\vec{R}_N$ according to $\langle R_N^2 \rangle \propto N^{2\nu}$ is $\nu=\frac{3}{4}$ in 2d. 
Note that when we increase the stiffness of the chain at fixed chain length, the times $\tau_1$ and $\tau_2$ can be made to coincide; 
this happens for $L=Nb=\ell_p$, as expected: when the contour length and the persistence length are of the same order, 
the regime described by Eqn.~\ref{gmid} is no longer present.
We will verify the time dependence of $g_1$ at various regimes from BD simulations.\par
We then obtain $\tau_2$ as follows:
\begin{equation}
g_1(t) = \ell_p^2 \left( t/\tau_1 \right) ^{3/5} \mathrm{~~~~~~for~~} t > \tau_1.
\end{equation}
Substituting $\tau_1$ from Eqn.~\ref{tau1} in above
\begin{equation}
g_1(t)  =  b^2 \left(\ell_p/b\right)^{1/5} \left( Wt\right)^{3/5},  \mathrm{~~for~~}  \tau_1 < t <  \tau_2.
\label{gmid}
\end{equation}
At $t = \tau_2$ 
\begin{equation}
g_1(t=\tau_2)  =   \langle R_N^2 \rangle = \ell_p^{1/2} L^{3/2}.
\end{equation}
Substituting Eqn.~\ref{gmid} for $t=\tau_2$ we get
\begin{equation}
W \tau_2  =   \left(\ell_p/b\right)^{\frac{1}{2}}N^{5/2}.
\label{tau2}
\end{equation}
We also note that the dynamics of the center of mass is given by 
\begin{equation}
g_3(t) = b^2 W \frac{t}{N}.
\label{g3_scaling}
\end{equation}
The ``phase diagram'' for the crossover dynamics in terms of $N$, and $\ell_p$ are shown in Fig.~\ref{phase}. 
Notice that for a stiffer chain the 
region for $\tau_1 < t < \tau_2$ for which we predict $g_1(t) \sim t^{0.6}$ requires to study very long chains and therefore, is  
hard to see in simulation for a stiffer chain. 

\section{THE MODEL}
\label{model_section}
\begin{figure}[ht!]                
\begin{center}
\includegraphics[width=0.9\columnwidth]{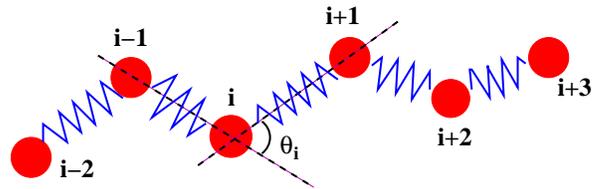}
\end{center}
\vskip -0.5truecm
\caption{\small Bead-spring model of a polymer chain with bending 
angle $\theta_i$ subtended by the vectors 
$\vec{b}_{i-1} = \vec{r}_{i}-\vec{r}_{i-1}$ and  $\vec{b}_{i} = \vec{r}_{i+1}-\vec{r}_{i}$.}
\label{model}
\end{figure}

We have used a bead spring model of a polymer chain with excluded volume, spring and 
bending potentials as follows~\cite{Grest_Kremer_PRA_1986}.
The excluded volume interaction between any two monomers is given by a short range Lennard-Jones (LJ) 
potential with cut off and shifted in its minimum.
\begin{eqnarray}
U_{\mathrm{LJ}}(r)&=&4\epsilon [{(\frac{\sigma}{r})}^{12}-{(\frac{\sigma}
{r})}^6]+\epsilon \;\;\;\;\;\;\;\; \mathrm{for~~} r\le 2^{1/6}\sigma  \nonumber \\
        &=& 0 \;\;\;\;\;\;\;\;\;\;\;\;\;\;\;\; \;\;\;\;\;\;\;\;\;\;\;\ \mathrm{for~~} r >  2^{1/6}\sigma\;.
\label{LJ}
\end{eqnarray}
Here, $\sigma$ is the effective diameter of a monomer, and
$\epsilon$ is the strength of the potential. The connectivity between
neighboring monomers is modeled as a Finitely Extensible Nonlinear Elastic (FENE) spring with 
\begin{equation}
U_{\mathrm{FENE}}(r)=-\frac{1}{2}kR_0^2\ln(1-r^2/R_0^2)\;,
\end{equation} 
where $r$ is the distance
between consecutive monomers, $k$ is the spring constant and $R_0$
is the maximum allowed separation between connected monomers~\cite{Grest_Kremer_PRA_1986}.
The chain stiffness is introduced by adding an angle dependent interaction between successive bonds
as (Fig.~\ref{model})
\begin{equation}
U_{\mathrm{bend}}(\theta_i) = \kappa(1-\cos \theta_i).
\label{ebend}
\end{equation}
Here $\theta_i$ is the complementary angle between the bond vectors 
$\vec{b}_{i-1} = \vec{r}_{i}-\vec{r}_{i-1}$ and 
 $\vec{b}_{i} = \vec{r}_{i+1}-\vec{r}_{i}$, respectively, as shown in Fig.~\ref{model}. The strength 
of the interaction is characterized by the bending rigidity $\kappa$. 
Introducing unit tangent vector $\mathbf{t}_i = \frac{\partial \mathbf{r}_i}{\partial s}$ ($|\mathbf{t}_i|=1$) we note that the 
discretized version of Eqn.~\ref{wlc} can be written as 
\begin{align}
\mathcal{H} \approx \frac{\kappa}{2}b_l \sum_i \left ( \frac{\mathbf{t}_{i+1}-  \mathbf{t}_i}{b_l} \right)^2
&=\frac{\kappa}{2b_l} \sum_i \left(\mathbf{t}_{i+1}^2 + \mathbf{t}_i^2 - 2\cos \theta_i \right) \nonumber \\
&=\frac{\kappa}{b_l}\sum_i\left( 1- \cos \theta_i \right), 
\end{align}
where $b_l=|\vec{b_i}|$ is the bond length in our simulation. Therefore, for fixed bond length Eqn.~\ref{ebend} represents  
the discrete version of the WLC model of Eqn.~\ref{wlc}.  
Thus the EV effect introduced through Eqn.~\ref{LJ} are completely absent in the WLC model.\par
We use the Langevin dynamics with the following equation of motion for the i$^{th}$ monomer 
\begin{equation*}
m \ddot{\vec{r}}_i=-\nabla (U_\mathrm{LJ}+U_\mathrm{FENE} + U_\mathrm{bend}) 
-\Gamma \vec{v}_i 
+ \vec{R}_i.
\end{equation*}
Here $\Gamma$ is the monomer friction coefficient and
$\vec{R} _ i (t)$, is a Gaussian white noise with zero mean at a temperature T, and
satisfies the fluctuation-dissipation relation:  
\[ < \, \vec{R} _ i (t)
\cdot \vec{R} _ j (t') \, > = 4k_BT \Gamma \, \delta _{ij} \, \delta (t
- t ')\;.\]
The reduced units of length, time and temperature are chosen to be  $\sigma$, 
$\sigma\sqrt{\frac{m}{\epsilon}}$, and $\epsilon/k_B$ respectively.  
For the spring potential we have chosen $k=30$ and $R_0=1.5\sigma$, the friction coefficient 
$\Gamma = 0.7$, the temperature is kept at $1.2/k_B$. The choice of the FENE potential along with the LJ interaction parameters 
ensures that the average bond-length in the bulk $\langle b_l \rangle = 0.971$. With the choice of these 
parameters the probability of chain crossing is very low. 
We also find that the average bond-length $\langle b_l \rangle$
is almost independent of the range of chain stiffness parameter ($\kappa = 0 - 320$) used in our simulation. 
Strictly speaking, the contour length $L$ is $L=\left(N-1\right)\langle b_l\rangle$.  
The equation of motion is integrated with the reduced time step $\Delta t = 0.01$ following the 
algorithm proposed by van Gunsteren and Berendsen~\cite{Langevin}.
\section{Results from Brownian Dynamics Simulation}
\label{BD_section}
We have carried out Brownian dynamics (BD) simulation for a wide range of chain length ($N$=16 - 2048) and 
bending constant ($\kappa$ = 0 - 320). Because of the argument given in Sec.~\ref{scaling_section} very long 
chains were needed to clearly identify the crossover regimes. First we present the equilibrium properties of the 
chains in section~\ref{eqlb_section} followed by the dynamical quantities presented in Sec.~\ref{dynamics_section}.
\subsection{Equilibrium Properties}
\label{eqlb_section}
\subsubsection{Persistence length}
From the BD simulation we have monitored the average $\langle \cos \theta \rangle$ 
and replacing $\cos \theta \rightarrow \langle \cos \theta \rangle$ in Eqn.~\ref{lp1} calculated 
the chain persistence length for various values of $\kappa$. 
One expects that Eqns.~\ref{lp2d} and \ref{lp1} must give results that agree with 
each other when the persistence length is much larger than the range of the excluded volume interaction, 
but that the two results agree even for small value of $\kappa = 2$ we believe is a nontrivial result. 
The comparison of calculated  $\ell_p$ by different methods is shown in Table~\ref{table}.
We also observe that the $\ell_p$ calculated using Eqn.~\ref{lp1} practically has no dependence on chain length $N$. 
We have used Eqn.~\ref{lp1} for further analysis of our data in the subsequent sections. Note that this behavior differs from the result found by 
Hsu {\em et al.} for a lattice model~\cite{Hsu_EPL_2011}, where a renormalization of $\ell_p$ by excluded volume was shown to occur. 
Thus on length scales of order $\ell_p$ there is no strict universality between different models.\par
We emphasize, however, that the persistence length, when it is supposed to measure the local intrinsic stiffness of the 
chain (as supposed in the Kratky-Porod model), cannot be estimated from the asymptotic decay of the bond vector correlation 
function $\langle \vec{b}_i \cdot \vec{b}_{i+s}\rangle$ with the ``chemical distance'' $s$ along the chain, that is the conventional 
definition given in all the polymer physics textbooks: as will be shown below (Sec.~\ref{bond_section}), we verify the predicted \cite{Hager} 
power law behavior for very long chains and large $s$, previously seen already for a lattice model of semi-flexible chains by 
Hsu {\em et al.}~\cite{Hsu_EPL_2011}. 
Although lattice and continuum models have different statistical properties when one considers lengths of the scale $\ell_p$, 
for much larger scales the behavior should be universal, and hence this power law decay is expected.
\begin{table}[htpb!]
\caption{Comparison of three ways of calculating $\ell_p$} \label{table}
\centering
\noindent\makebox[\columnwidth]{
\begin{tabular}{| c || c || c || c | }  \hline
  ~~$\kappa$~~ &  Eqn.~\ref{lp2d}   &  Eqn.~\ref{lp1} & $\ell_p$ from fitted slope (Fig.~\ref{semilog})\\
\hline
64 &     106.7     &        105.8  & 112.4 \\
\hline
32 &     53.3      &        52.6   &  53.7 \\
\hline
16 &     26.7   &       25.9      &  27.4  \\
\hline
8  &     13.3     &       12.6    &  13.7  \\
\hline
4  &     6.7       &       6.05    &  6.7  \\
\hline
2  &     3.3      &       3.31    & 4.2   \\ \hline
\end{tabular}}
\end{table}
We also note that a definition of the persistence length dating back to Flory, where one considers the correlation of the 
first bond vector $\vec{b}_1$ with the end-to-end vector $\vec{R}_N$,
\begin{equation}
\ell_p^{\textrm{Flory}} = \langle \vec{b}_1 \cdot \vec{R}_N\rangle /\langle b \rangle \;,
\end{equation}
which has been advocated by Cifra~\cite{Cifra} as an ``exact expression'', must similarly be refuted: in 2d, Redner and 
Privman~\cite{Redner} have shown that $\ell_p^{\textrm{Flory}}$ for large $N$ is logarithmically divergent with $N$ already 
for a simple self-avoiding walk (SAW). For completeness, we mention that an analogous definition for inner bond vectors
\begin{equation}
\ell'_p = \langle \vec{b}_i \cdot \vec{R}_N\rangle /\langle b \rangle \quad , \quad 1 \ll i \ll N
\end{equation}
even shows a power-law divergence, $\ell_p' \propto N^{(2 \nu -1)} $, 
both in 2d and 3d (see Hsu et al.~\cite{Hsu_Macro_2010} and \cite{Elsner}). Thus we urge that the results and discussion presented in 
this section need to be taken 
seriously in writing future review articles and newer edition of the existing textbooks. 
\subsubsection{Scaling of semi-flexible chain; comparison with theory}
\label{scaling_rn}
\begin{figure}[ht!]                
\begin{center}
\includegraphics[width=0.9\columnwidth]{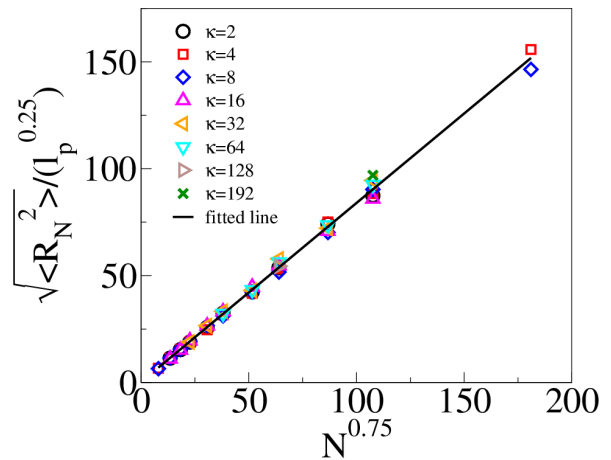}
\end{center}
\vskip -0.5truecm
\caption{\small Plot of $\sqrt{\langle R_N^2 \rangle}/\ell_p^{0.25}$ versus $N^{0.75}$ for 
various values of the chain stiffness parameter. All the data for different stiffness parameter collapse 
on the same master plot. The solid line is a fit to a straight line. Only data points for which the contour length exceeds 
the persistence length were included in this plot.} 
\label{r1n}
\end{figure}

The extension of Flory theory for a semi-flexible chain has been done by Schaefer, Joanny, and Pincus~\cite{Pincus_MM_1980} 
and Nakanishi~\cite{Nakanishi_1987} which states that the RMS of the end-to-end distance $\langle R_N^2 \rangle$ in $d$ spatial dimensions 
exhibits the following scaling relation
\begin{equation}
\sqrt {\langle R_N^2 \rangle} \sim N^{\frac{3}{d+2}}\ell_p^{\frac{1}{d+2}}b^{\frac{d+1}{d+2}},
\label{scaling}
\end{equation}
where $b$ is the bond length ($\langle b_l \rangle$ in our simulation). For $d=2$ this reduces to 
$\sqrt {\langle R_N^2 \rangle} \sim N^{0.75}\ell_p^{0.25}$. 
In other words if the end-to-end distance is scaled by the appropriate power of the persistence length $\ell_p$, then this renormalized 
end-to-end distances $\langle \tilde{R}_N \rangle  = \sqrt{\langle R_N^2 \rangle}/\ell_p^{0.25}$ for different values of the chain stiffness parameter 
$\kappa$ will fall onto the same
master plot. For a large combination of chain length $N$ and stiffness parameter $\kappa$ we observe excellent fit to our equilibrium 
data for $ \langle \tilde{R}_N \rangle $ to Eqn.~\ref{scaling} as shown in Fig.~\ref{r1n}. 
It is worth noting that the persistence length calculated from 
Eqn.~\ref{lp1} using the formula from WLC model uses the local correlation, namely the angle between the subsequent bond vectors 
and hence is expected to provide a decent value of the persistence length when EV is also included. 
The excellent collapse of the data for $\tilde{R}_N \sim N^{0.75}$ for various 
values of the stiffness parameter ($\kappa = 1 - 192$) on the same master curve indicates that Eqn.~\ref{lp1} {\em can be used as the standard 
definition of persistence length even in presence of the EV effect}.
\subsubsection{Comparison with WLC in 2d}
Having established the definition of persistence length which validated Eqn.~\ref{scaling}, 
we now use Eqn.~\ref{rn_wlc} presented in section~\ref{WLC_section} to analyze the BD simulation results for the end-to-end distance. 
Please note that limiting cases of Eqn.~\ref{rn_wlc} are either a Gaussian coil ($\langle R_N^2\rangle = 2\ell_pL$ for $ L \gg \ell_p $) 
or a rod ($\langle R_N^2\rangle = L^2$ for $ L \ll \ell_p $). We have used simulation results to 
plot $\frac{\langle R_N^2\rangle}{2\ell_pL} \sim L/\ell_p $  for a large number of values($\sim 100$) of $L/\ell_p$ ($0.05 \le L/\ell_p \le 170$).
\begin{figure}[ht!]                
\begin{center}
\includegraphics[width=0.9\columnwidth]{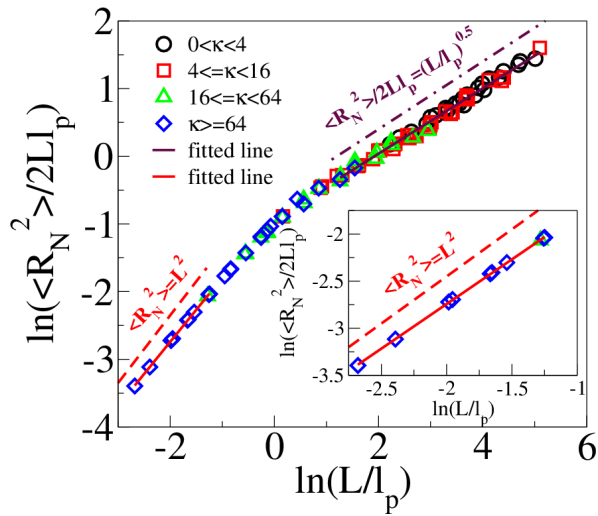}
\end{center}
\vskip -0.5truecm
\caption{\small $\langle \langle R_N^2\rangle/(2L\ell_p)$ as a function of $L/\ell_p$ obtained from 
different combinations of chain length $N$ and stiffness parameter $\kappa$ (log-log scale). The solid (maroon) line 
is a fit to the formula $\langle R_N^2\rangle/2L\ell_p \sim (L/\ell_p)^{0.5}$ for $4 < L/\ell_p < 170$.
The inset shows the same for small values of $0 < L/\ell_p< 1$ which clearly indicates that limiting 
slope of unity ($\langle R_N^2 \rangle = L^2$) for $L/\ell_p \rightarrow 0$. }
\label{rn}
\end{figure}
We have also taken additional care that a given value of $L/\ell_p$ is generated for different combinations of $L$ and $\ell_p$. These results 
are shown in Fig.~\ref{rn}. For $L/\ell_p \ll 1$ we observe that $\frac{\langle R_N^2\rangle}{2\ell_pL} \sim (L/\ell_p)^{0.95}$ while for 
$L/\ell_p \gg 1$ the data very nicely fit with $\frac{\langle R_N^2\rangle}{2\ell_pL} \sim (L/\ell_p)^{0.50}$. This is consistent with prior MC results 
using a lattice model by Hsu {\em et al.}~\cite{Hsu_EPL_2011}. However, since our studies are done in continuum we are able to get data that is for much shorter 
length scales. The fact that in the rod-like regime ($L<\ell_p$) the ``best fit" 
exponent is 0.95 rather than the asymptotic value 1.0 is due to the fact that for $\kappa$=32 and 64 the ``rods" still 
exhibit nonnegligible transverse fluctuation unlike truly stiff rods. The Gaussian behavior that Eqn.~\ref{rn_wlc} implies $\langle R_N^2 \rangle/2L\ell_p=1$ for 
large $L$ would mean a horizontal straight line in Fig.~\ref{rn}, but no indication of such a behavior is seen. The simulation data then implies 
the strict absence of a Gaussian limit for 2d swollen semi-flexible chains due to severe dominance of the EV interaction. This result 
should be contrasted with the simulation results in 3d, where one sees a gradual 
crossover from rod limit to the EV limit (in 3d) passing through a Gaussian regime~\cite{Moon_1991,Hsu_EPL_2010}. 
\subsubsection{Transverse Fluctuations}
It is reasonable to define an average axis for a polymer chain in the rod limit ($\ell_p \gg L$). In this limit 
using WLC chain Hamiltonian the transverse fluctuation with respect to this average axis has been shown~\cite{Fuji,Odijk,Capsi} to obey 
the following equation
\begin{equation}
\langle l_{\bot}^2\rangle \sim L^3/\ell_p , \label{transEq}.
\end{equation}
The above equation implies that the roughness exponent~\cite{Barabasi} $\zeta=3/2$ ($ \sqrt{\langle l_{\bot}^2\rangle} \sim L^\zeta$) 
for a weakly bending rod. Starting from an extremely stiff chain where the transverse fluctuations are expected 
to be governed by the roughening 
exponent $\zeta=3/2$, if we approach the limit of fully flexible chain, then the fluctuations become isotropic and 
in this limit one can expect that $\langle l_{\bot}^2\rangle \sim L^\nu$, so that in 
2d $\langle l_{\bot}^2\rangle \sim L^{0.75}$. In order to calculate the transverse fluctuation, 
in our simulation, for each configuration of the polymer chain, we choose $\hat{x}=\vec{R}_N/R_N$ as the 
longitudinal axis and calculate transverse fluctuations as follows:  
\begin{equation}
\langle l_{\bot}^2\rangle = \frac{1}{N}\sum_{i=1}^N y_i^2, \label{lbot}
\end{equation}
where $y_i$ is the perpendicular distance of the $i^{th}$ monomer with respect to the instantaneous direction $\vec{R}_N$.
We have repeated this calculation for several chain lengths from extremely stiff chains to fully flexible chains. 
\begin{figure*}[htbp]
\centering
\includegraphics[height=6.0cm]{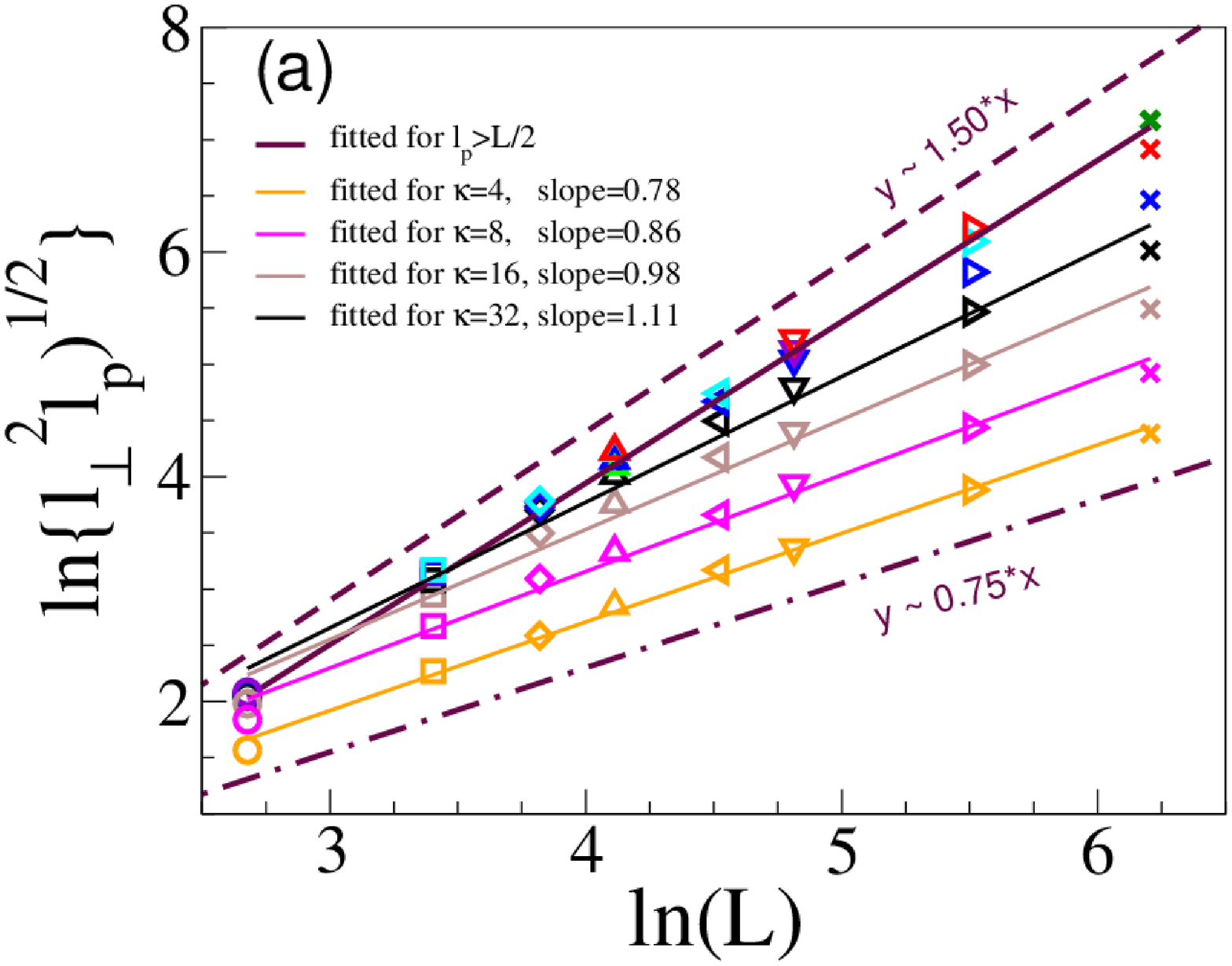}
\includegraphics[height=6.0cm]{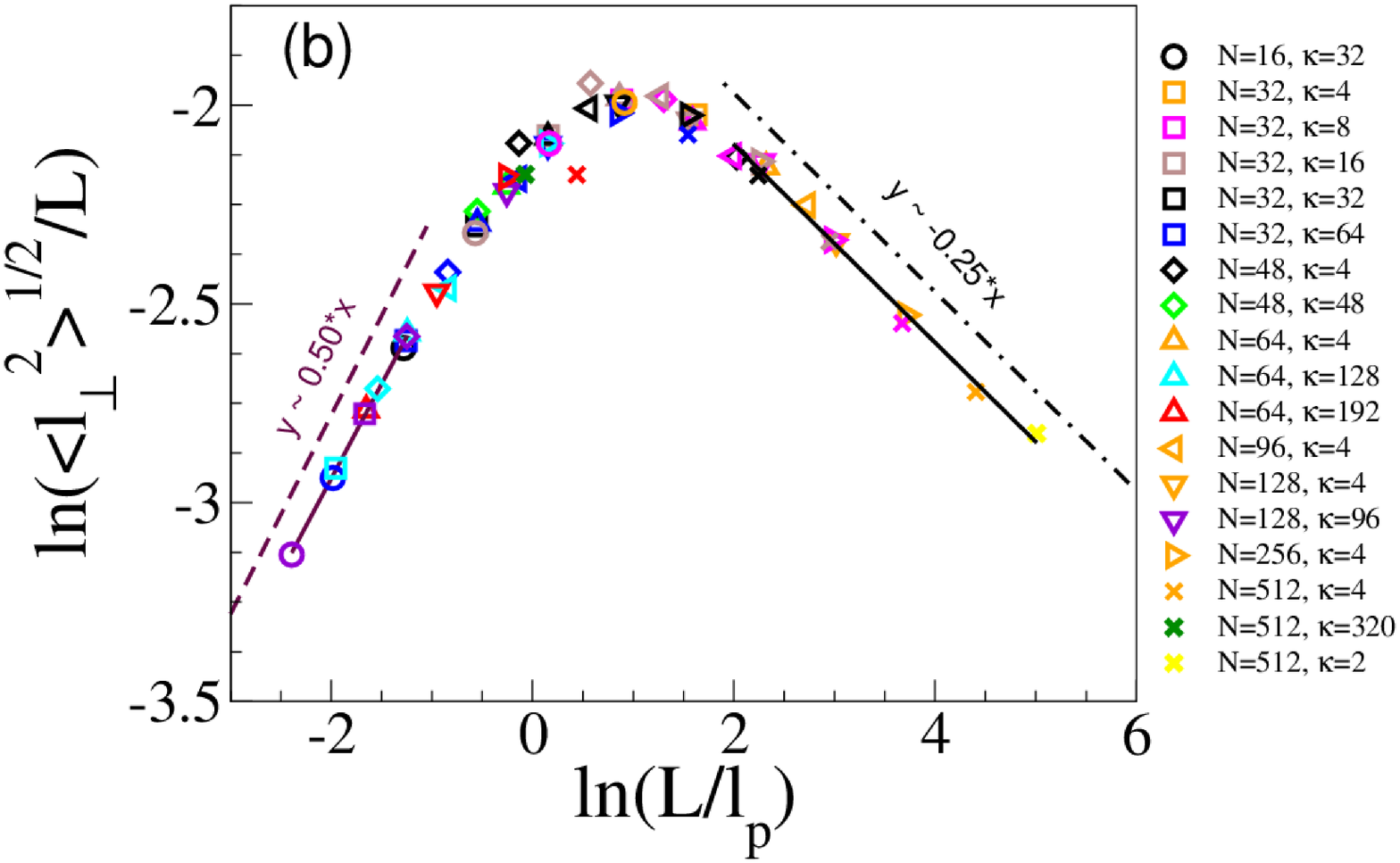}
\caption{(a) Log-log plot of $\sqrt{\langle l_{\bot}^2\rangle \ell_p}$ as a function of the contour length $L$ for chains of chain length 
$N=16$ (circle), 32 (square), 48 (diamond), 64 (up triangle), 96 (left triangle), 128 (down triangle), 256 (right triangle), 512($\times$) 
with various bending stiffness $\kappa=2$ (yellow), 4 (orange), 8 (magenta), 16 (brown), 32 (black), 48 (green), 64 (blue), 96 (indigo), 128 (cyan), 
192 (red), 320 (dark green). The orange, magenta, brown, black solid lines are fitted to chains with $\kappa=4, 8, 16, 32$, respectively.
The maroon solid line is fitted for chains 
with $\ell_p>L/2$. (b) Log-log plot of $\sqrt{\langle l_{\bot}^2\rangle}/L$ as a function of $L/\ell_p$ showing excellent data collapse as well as 
maxima of the rescaled fluctuations around corresponding rescaled length $L/\ell_p \approx 1$.}
\label{tfl}
\end{figure*}
This is shown in Fig.~\ref{tfl}(a). 
If one does not analyze the data carefully, one might be misled to conclude that the exponent $\zeta$ increases gradually with the 
chain stiffness $\kappa$. However, as we will see that the proper interpretation requires a scaling description in terms of $L/\ell_p$ as scaling variable.
The interesting aspect of this rescaled dimensionless transverse fluctuation is shown in Fig.~\ref{tfl}(b) as a 
function of rescaled length $L/\ell_p$ where the rescaled fluctuations collapse on the same master curve. 
This plot exhibits a maximum and then decreases gradually for large value of $L/\ell_p$. 
It is worth noting that analytical results do not exist for chains with intermediate stiffness.
The physical origin of this maximum can be interpreted as follows. Starting from the stiff chain limit when $L/\ell_p \ll 1$   
it increases for a more flexible chain when the chain undergoes shape changes from rod to ellipsoid-like blob, and finally to isotropic spherical blobs.
Naturally, when chain flexibility is defined in units of the persistence length for some value of  $L/\ell_p$ ($L/\ell_p \sim$ 3 from Fig.~\ref{tfl}(b)) 
the fluctuation becomes maximum before it becomes 
isotropic. To the best of our knowledge this result is new and in principle can be 
used to measure the persistence length of a semiflexible polymer by measuring the transverse fluctuations
using fluorescence probes. This would require numerically analyzing a large number of images of semiflexible chains (of a given kind of polymer) with varying contour length. For each image of a chain one can extract $L$ as well as the end-to-end vector $\vec{R}_N$ and then use Eqn.~\ref{lbot} to extract $\langle l_{\bot}^2\rangle$. Plotting then $\sqrt{\langle l_{\bot}^2\rangle}/L$ versus $L$ one would find $l_p$ from the position of the maximum of this plot.
\subsubsection{Bond vector correlation}\label{bond_section}
The orientational correlation between successive bonds decays exponentially in a WLC model according to Eqn.~\ref{bexp}.
However, recent Monte Carlo (MC) studies by Hsu, Paul, and Binder have verified that a swollen semiflexible chain in 2d exhibits a power law 
decay as a function of the separation $s$ between the beads~\cite{Hsu_EPL_2011} given by  
\begin{equation}
\left< \hat{b}_i \cdot \hat{b}_{i+s}\right> \propto s^{-\beta},\; \beta=2-2\nu \;\; {\rm for}\; s \ll N.
\label{bpower}
\end{equation}
For a fully flexible chain $\nu = 0.75$ in 2d so that $\beta = 0.5$. It is then expected that a semiflexible chain will exhibit 
the same behavior when $L/\ell_p \gg 1$. While for very stiff chain one needs to have extremely long chain to see this asymptotic regime
for large $L/\ell_p \gg 1$, yet satisfying the condition  $s \ll N$, from our simulation we clearly see this trend for moderate values of 
$\kappa$.  \par
\begin{figure}[htbp!]
\centering
\includegraphics[width=0.9\columnwidth]{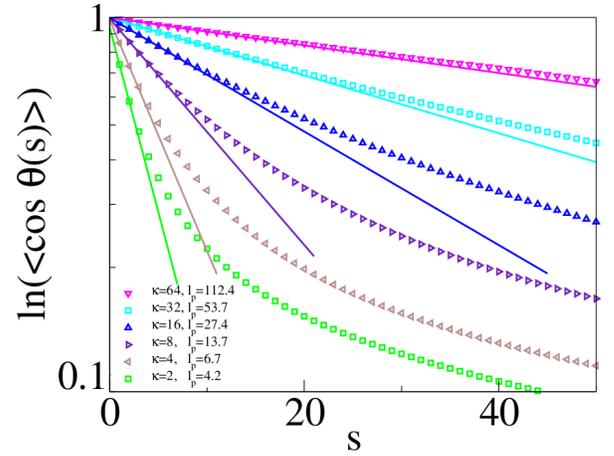}
\caption{\small Semi-log plot. $\ell_p$ is calculated from the slope of the fitted line, which is very close to 
the values (correspondingly 105.8, 52.6, 25.9, 12.6, 6.05, 3.31) from 
$\ell_p=-1/\mathrm{ln} \langle \mathrm{cos}(\theta) \rangle$ (see Table~\ref{table}). }\label{semilog}
\end{figure}

 
We calculated the bond correlation function from its definition and tested both Eqn.~\ref{bexp} and \ref{bpower}. 
First in Fig.~\ref{semilog} we show the semi-log plot of Eqn.~\ref{bexp}. 
The straight lines are fitted only with the first several data points in 
order to get values of $\ell_p$ which are close to those calculated from $\ell_p=-1/\mathrm{ln} \langle \mathrm{cos}(\theta) \rangle$. 
The deviation from the initial slope (only after few points) clearly shows that Eqn.~\ref{bexp} predicted by the WLC model does not hold 
good for a 2d swollen chain as expected from the result of Fig.~\ref{rn}.  \par
\begin{figure}[htbp!]
\centering
\includegraphics[width=0.94\columnwidth]{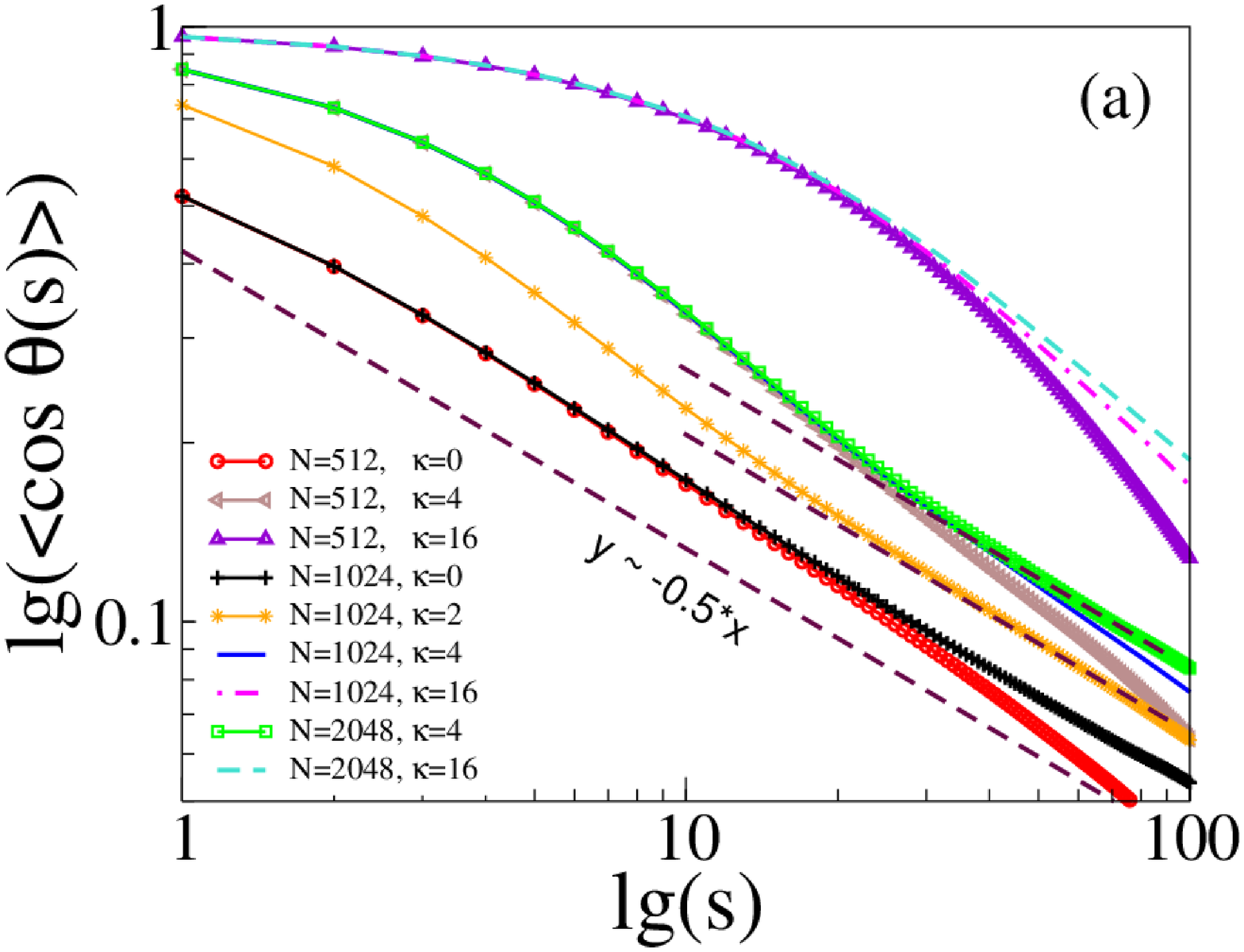}\\
\includegraphics[width=0.9\columnwidth]{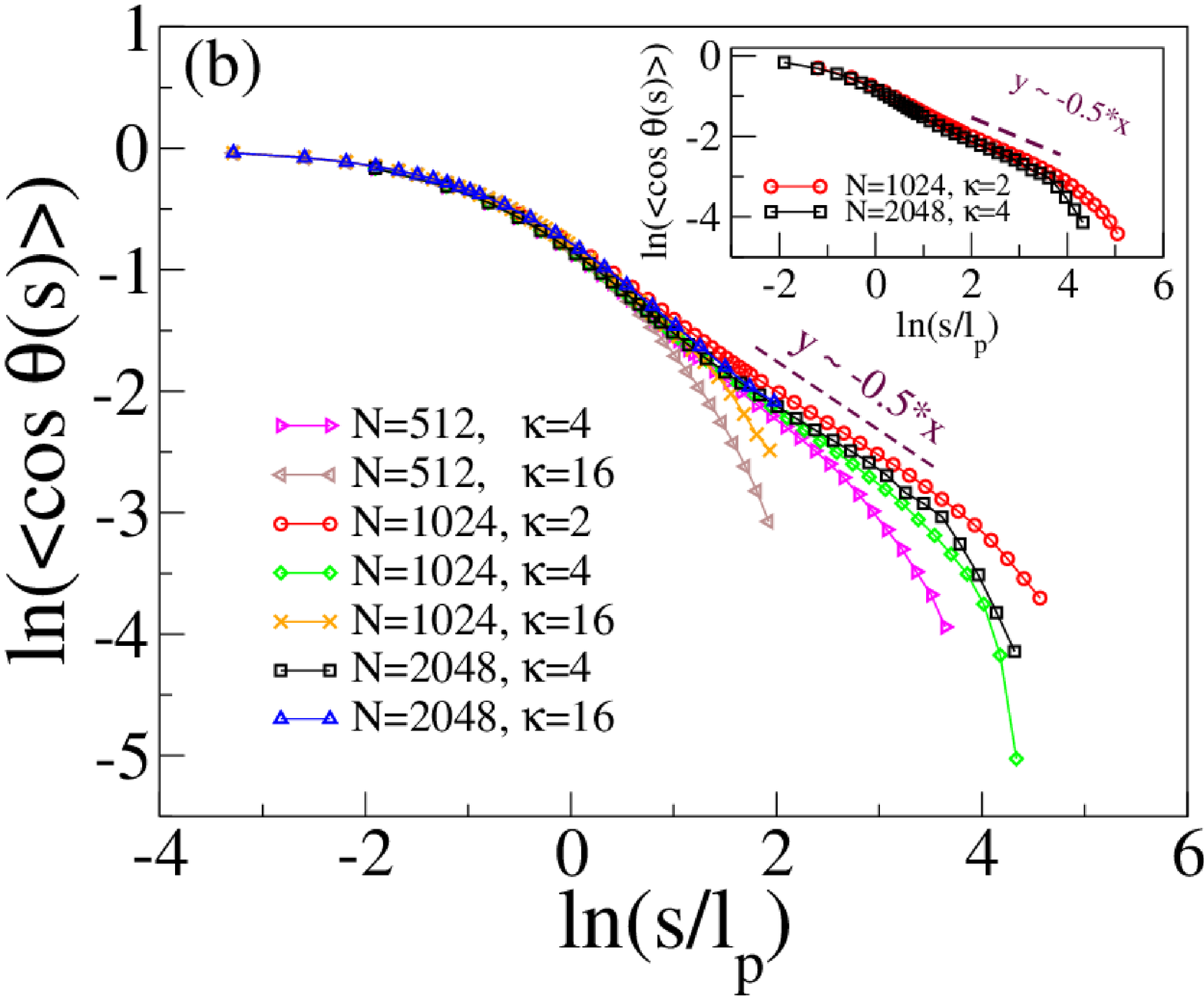}
\caption{\small (a) Log-log plot of $\langle \cos \theta(s) \rangle$ as a function of $s$ 
for various combinations of $N$ and $\kappa$. While for large $\kappa$ the asymptotic slope 
of $s^{-0.5}$ is preempted by finite size effect, for $(N,\kappa) \equiv (1024,2)$ and 
$(N,\kappa) \equiv (2024,4)$ the slope of $-0.5$ is clearly visible; (b) Same 
as in (a) but using rescaled variable $s/\ell_p$ which shows excellent data collapse for $s/\ell_p \le 1$. The inset shows 
the same but only for the cases to emphasize that in the limit $L/\ell_p \gg 1.0 $ there is a perfect 
data collapse for the power law scaling. }\label{loglog}
\end{figure}

 
Fig.~\ref{loglog} shows the log-log plot of Eqn.~\ref{bpower} where we have also included the graph for a fully flexible chain for reference. 
Please note that even for a fully flexible chain it requires a rather long chain length ($N=1024$) to clearly see the regime with slope with 
$\beta=0.5$ over an appreciably broad range of abscissa values. For comparison we put the graph for a shorter fully flexible chain of $N=512$. Here the curve starts to decrease faster before it reaches the slope corresponding to $\beta=0.5$. Naturally for stiffer chains
(which could be thought of a flexible chain of length $L/\ell_p$ for this purpose) for the maximum chain length $N =2048$ considered in this paper we 
only see the regime characterized by $\beta = 0.5$ only for $\kappa=2.0$ and 4.0. \par
It is expected that the bond vector correlation also exhibits a scaling behavior when studied as a function of $s/\ell_p$, as 
even for moderate values of $\ell_p \sim 5$ (see Fig.~\ref{semilog}), 
this length rescaling $L \rightarrow L/\ell_p$ overrides the exponential decay for 
small $s/\ell_p$. With a choice of distance between monomers in units of $\ell_p$ all chains are expected to behave as fully flexible chains.   
Therefore, if we use 
the renormalized distance $s/\ell_p$ to replot Fig~\ref{loglog}(a) then one expects that the power law correlation for chains with 
different stiffness will collapse on the same master plot. This is shown in Fig.~\ref{loglog}(b). 
We observe excellent data collapse for $s/\ell_p \le 1$. Deviations from this collapse occur at a progressively larger value of $s/\ell_p$ as the 
ratio $L/\ell_p$ increases either by increasing the contour length $L$ for a fixed $\ell_p$ or for the same contour length $L$ and lowering the value of $\ell_p$. 
This is expected, since Eqn.~\ref{bpower} can hold only for $s \ll N$. 
Of course, a numerical study of the regime $\ell_p/\langle b \rangle \ll s \ll N$ for large $\ell_p$ is prohibitively difficult. 
However, the inset shows both the data collapse and the $\beta = 0.5$ regimes for two chain lengths ($N=1024$ and 2048) and for 
two values of $\kappa$ ($\kappa = 2$ and $\kappa=4$) which proves beyond 
doubt that for $1 \ll s/\ell_p \ll L/\ell_p$ the bond autocorrelation exhibits a power law decay of a fully flexible chain.   
\subsubsection{Radial distribution function for end-to-end distance}
\begin{figure*}[htbp]
\centering
\includegraphics[height=8.0cm]{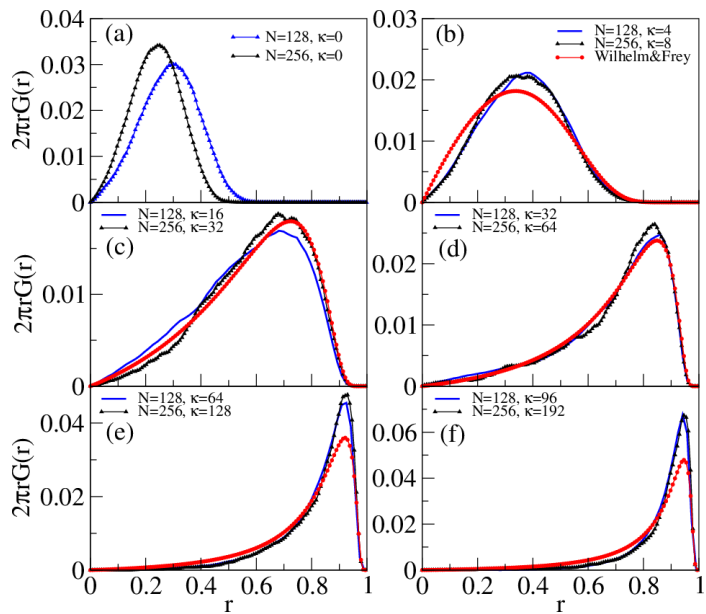}
\caption{ Comparison of the radial distribution for Eqn.~\ref{Freyl0} and the simulation for N=128 and 256 
with various values of $\kappa$.}\label{distri}
\end{figure*}
The Hamiltonian for the WLC chain model has been studied by many analytic technics assuming that for moderate chain lengths and stiff enough 
chains the EV effect will not dominate. Since we already established the severe dominance of the EV effect, Gaussian regime is absent for 
a 2d swollen chain. Here we compare the radial distribution functions for chains with different stiffness. In particular, we compare the results 
from our simulation with analytic results of Wilhelm and Frey~\cite{Wilhelm_PRL_1996} in 2d.
For the WLC model Wilhelm and Frey~\cite{Wilhelm_PRL_1996} have derived expressions for the radial distribution functions $G(r)$ (both in 2d and 3d) in terms of infinite series. 
In 2d the expression for $G(r)$ is given as follows.
\begin{align}
G(r) = \mathrm{const} \times t \sum^{\infty}_{l=0} \frac{(2l-1)\,!!}{2^ll\,!} \left[ \frac{1}{2t(1-r)} \right]^{5/4} \times \nonumber \\
\mathrm{exp}\left[ -\frac{(l+1/4)^2}{2t(1-r)}\right] D_{3/2}\left[ \frac{2l+1/2}{\sqrt{2t(1-r)}} \right], \label{Frey}
\end{align}
where $t=\ell_p/L$, $r=R_N/L$ and $D_{3/2}(x)$ is a parabolic cylinder function. For the whole range of values of $\kappa$ in our simulation, 
when we plot the analytic result of Eqn.~\ref{Frey} we find that 
this series is fully dominated by the first term and we didn't see visual differences from the plot that 
includes the first four terms. 
Therefore, in Fig.~\ref{distri} we plot only the first term of this series  ($l=0$) term given by 
\begin{align}
G_{l=0}(r) = \mathrm{const} \times t \times \left[ \frac{1}{2t(1-r)} \right]^{5/4} \times \nonumber \\
\mathrm{exp}\left[ -\frac{(1/4)^2}{2t(1-r)}\right] D_{3/2}\left[ \frac{1/2}{\sqrt{2t(1-r)}} \right] \label{Freyl0}
\end{align}
Please note that for the radial distribution function Wilhelm and 
Frey~\cite{Wilhelm_PRL_1996} does not take into account excluded volume effects.
In Fig.~\ref{distri} we have also included the distribution for a completely flexible chain ($\kappa=0$) for comparison purposes. For 
$\kappa=4$ and 8 ({\em i.e.}, $L/\ell_p \approx 16$) the simulated distribution still is intermediate between the behavior of 
fully flexible chains (which we expect to result for $N\rightarrow \infty$ for all $\kappa$!) and the distribution of the 
Kratky-Porod model. 
\subsection{Dynamics}\label{dynamics_section}
We now look at the dynamics of a swollen semi-flexible chain.  We are using a
Brownian dynamics (BD) scheme and therefore, HD effects are not included in our studies. 
However, for polymers adsorbed on a flat surface simulation studies have shown that the HD effects are negligible for no-slip boundary conditions~\cite{Winkler}. Thus we expect the results are 
relevant for a fair comparison of  experimental studies. 
Also we would like to point out that unlike equilibrium properties, 
computational time increases dramatically
to study time dependent properties for the same chain length. 
Thus the diffusion and relaxation studies for the longest chain 
length reported in this paper took significant time for well converged runs. 
We verified that the diffusion constant $D$ is independent of the persistence length $\ell_p$ and depends only on the chain length $N$ and 
scales as $D \sim k_BT/N$  with very good accuracy, as expected in a BD formalism.\par
Next we consider monomer relaxation dynamics of the chain. 
Following previous work 
for the relaxation of a fully flexible chain ~\cite{Grest_Kremer_PRA_1986,Gerroff_JCP_1992,Milchev_JCP_1993,Binder_Review} 
we have monitored various quantities 
pertaining to a single monomer relaxation. These quantities have been studied in the past using 
BD algorithm~\cite{Grest_Kremer_PRA_1986}, 
and dynamical Monte Carlo algorithms (DMC)~\cite{Gerroff_JCP_1992,Milchev_JCP_1993}, including the  bond-fluctuation model (BFM)~\cite{Binder_Review}. 
The dynamics of the individual monomers and the collective dynamics for the whole chain have 
been characterized by the functions $g_1(t)$, $g_2(t)$, $g_3(t)$, $g_4(t)$, and $g_5(t)$. They are defined as follows:
\begin{subequations}
\begin{align}
&g_1(t) = \langle \left[ \mathbf{r}_{N/2}(t)  - \mathbf{r}_{N/2}(0)\right]^2\rangle, \label{g1}\\
&g_2(t) = \nonumber \\ 
&\langle \left[ \left(\mathbf{r}_{N/2}(t)  - \mathbf{r}_{CM}(t) \right)  -  \left( \mathbf{r}_{N/2}(0)  - \mathbf{r}_{CM}(0)\right) \right]^2 \rangle,\label{g2}\\
&g_3(t) = \langle \left[ \mathbf{r}_{CM}(t)  - \mathbf{r}_{CM}(0)\right]^2 \rangle, \label{g3}\\
&g_4(t) = \langle \left[ \mathbf{r}_{end}(t)  - \mathbf{r}_{end}(0)\right]^2 \rangle, \label{g4}\\
&g_5(t) = \nonumber \\
&\langle \left[ \left( \mathbf{r}_{end}(t)  - \mathbf{r}_{CM}(t)\right)  -    
\left(\mathbf{r}_{end}(0)  - \mathbf{r}_{CM}(0)\right) \right]^2 \rangle. \label{g5}
\end{align}
\end{subequations}
The quantities $g_1(t)$, $g_3(t)$, and $g_4(t)$ reflect the time dependence of the position of the middle monomer, the center of mass, 
and the end monomer of the chain respectively.  
At late times for distances greater than the gyration radius the functions $g_1(t)$, $g_3(t)$ and $g_4(t)$ will 
describe the motion of the entire chain. Consequently, 
\begin{equation}
g_1(\tau(N,\ell_p)) \sim g_3(\tau(N,\ell_p)  \sim g_4(\tau(N,\ell_p) \sim t ,
\end{equation}
for $t  \geq \tau_2(N,\ell_p)$ where $\tau_2$ is given by Eqn.~\ref{tau2}. 
The quantities $g_2(t)$ and $g_5(t)$ on the contrary measure the relative displacement of the middle and the end monomer with 
respect to the center of mass of the chain. The functions $g_2(t)$ and $g_5(t)$ saturate at finite static values 
$2\langle \left( \vec{r}_{N/2}-\vec{r}_{CM}\right)^2\rangle$ and  $2\langle \left( \vec{r}_{end}-\vec{r}_{CM}\right)^2\rangle$ respectively, 
since for $t\rightarrow \infty$ the orientations of the vectors $\vec{r}_{N/2}(t)-\vec{r}_{CM}(t)$, $\vec{r}_{end}(t)-\vec{r}_{CM}(t)$ 
are uncorrelated with their counterparts at $t=0$.\par
\begin{figure}[ht!]
\begin{center}
\includegraphics[width=\columnwidth]{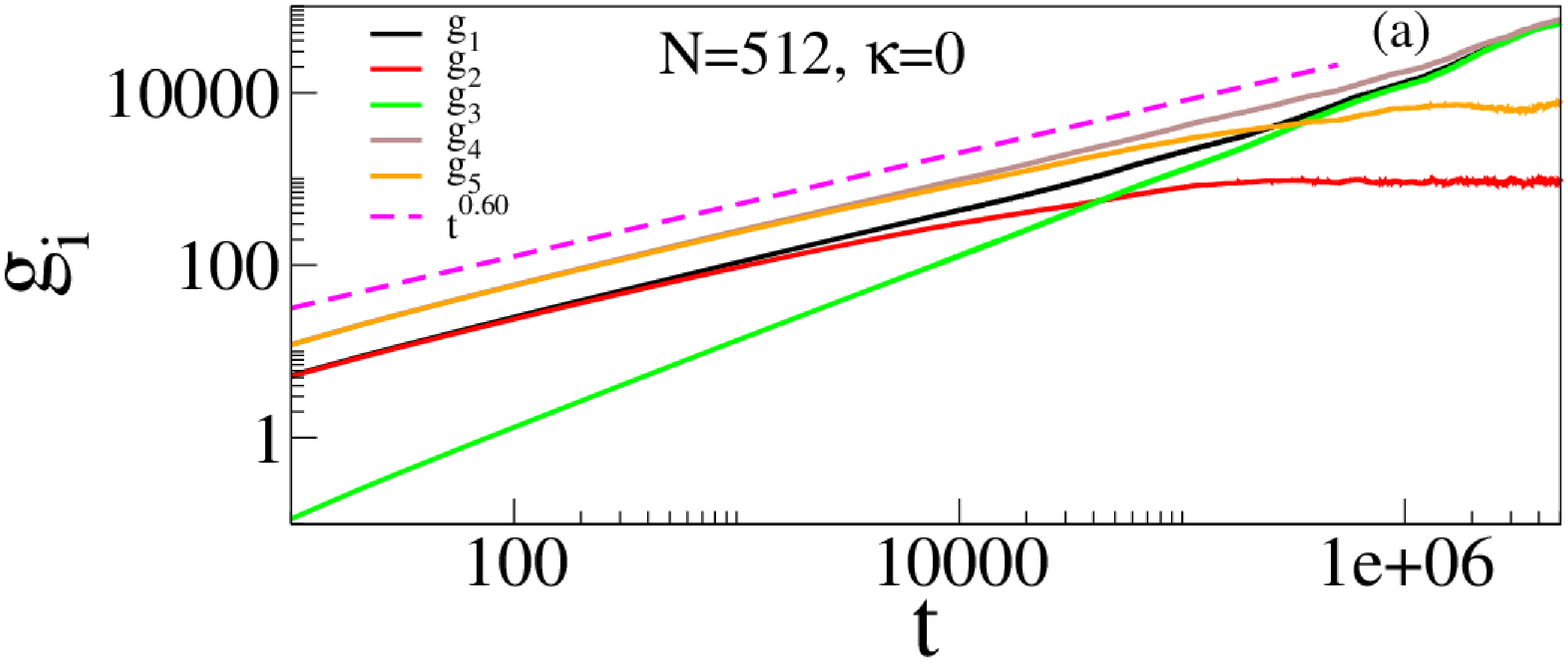}\\
\includegraphics[width=1.01\columnwidth]{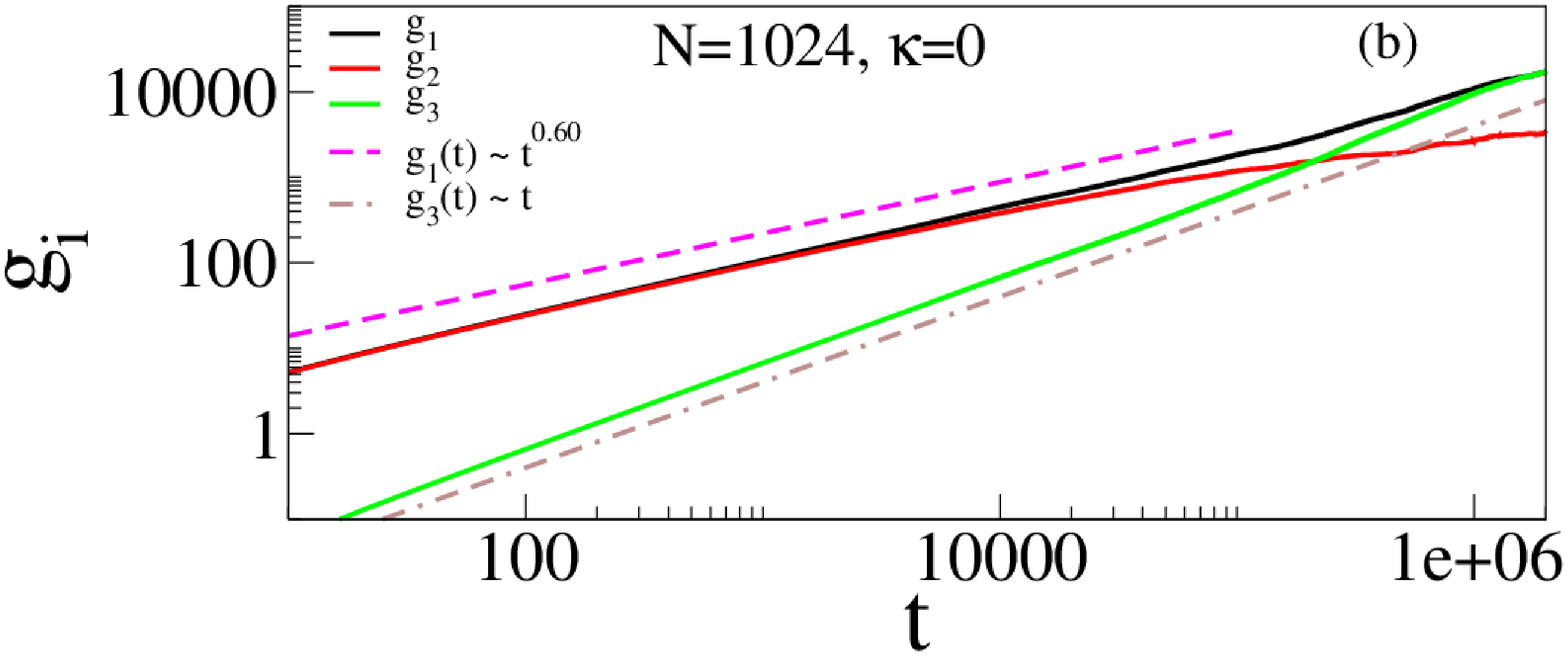}
\caption{\small (a) $g_i(t) \sim t$ ($i=1-5$) for a fully flexible chain ($\kappa = 0.0$) of length $N=512$. (b) 
$g_i(t) \sim t$ ($i=1-3$) for a fully flexible chain of length $N=1024$.}
\label{kappa0}
\end{center}
\end{figure}

To study monomer dynamics, we have carried out BD simulation for various chain lengths $N=64-1024$ and for chain stiffness $\kappa = 0 - 64$. 
We only show a limited set of data. As a reference and for comparison with the data for chains with $\kappa \ne 0$, we first show data for 
a fully flexible chain where we expect to see a single crossover dynamics from $g_1(t) \sim t^{0.6}$ at an early time for $ 0 < t < \tau_2$ to a 
purely diffusive dynamics for the entire chain ($g_1(t) \sim t$).   
This is shown in Fig.~\ref{kappa0} for chain length $N=512$ and 1024 respectively. 
We have checked that the graphs for other chain length are similar and 
have the $t^{0.6}$ dependence for the fully flexible chain. 
At late times the functions $g_2$ and $g_5$ saturates and the functions $g_1$, $g_3$ and $g_4$ grow linearly as a function of time, becoming 
practically indistinguishable from each other. Similar studies  
have been reported earlier by Grest and Kremer~\cite{Grest_Kremer_PRA_1986} and by 
Gerroff, Milchev, Paul, and Binder~\cite{Gerroff_JCP_1992,Milchev_JCP_1993,Binder_Review}. 
However, our studies are much more exhaustive and quantitatively  
captures the crossover from $t^{0.6}$ to the purely diffusive regime which were only qualitative in previous studies and for shorter chains.
\par
\begin{figure}[ht!]                
\centering
\includegraphics[width=\columnwidth]{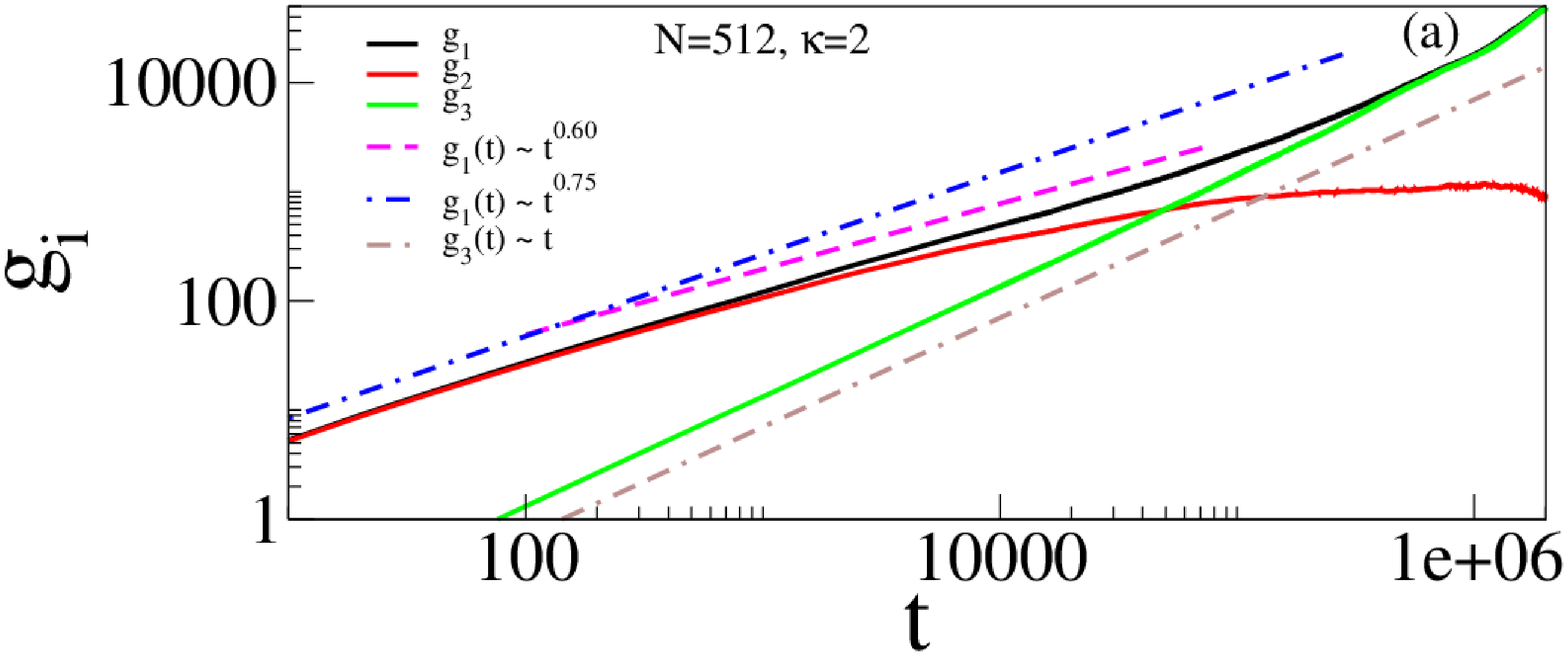}\\
\includegraphics[width=\columnwidth]{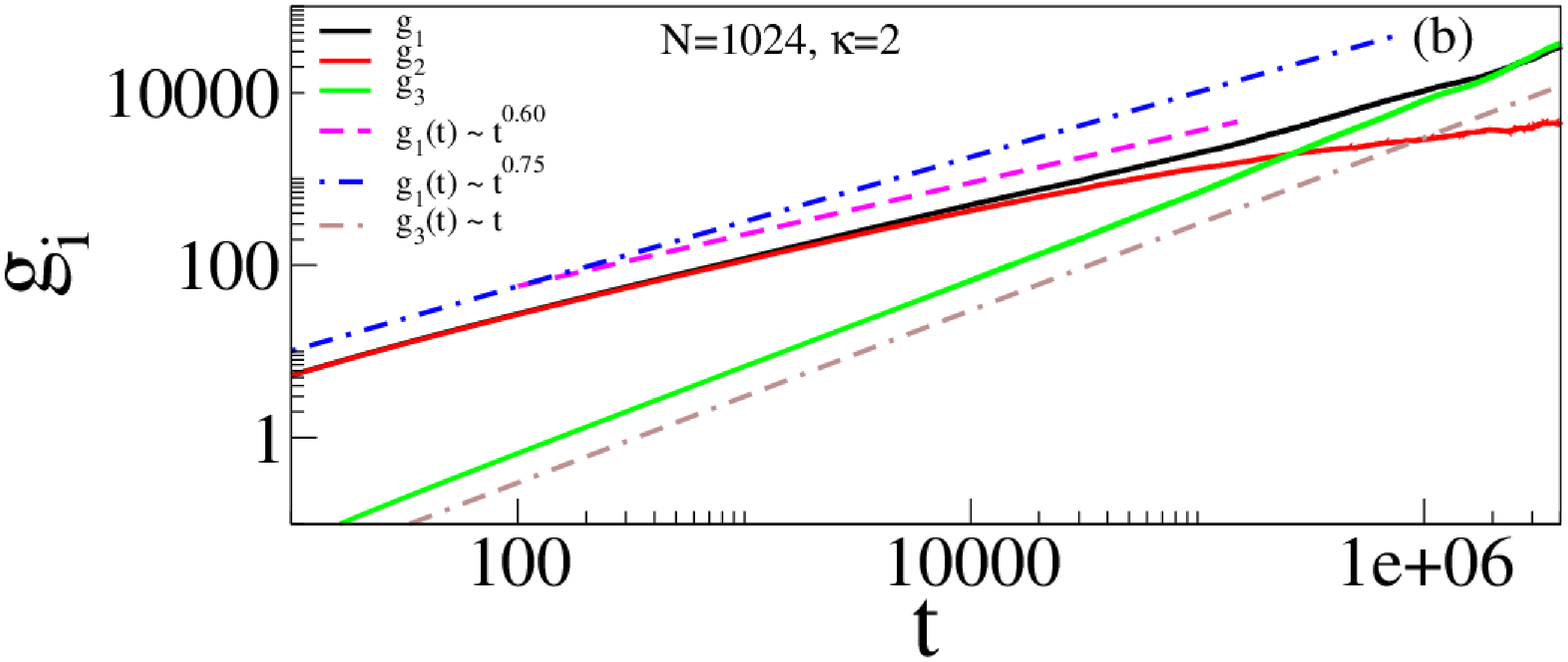}\\
\includegraphics[width=\columnwidth]{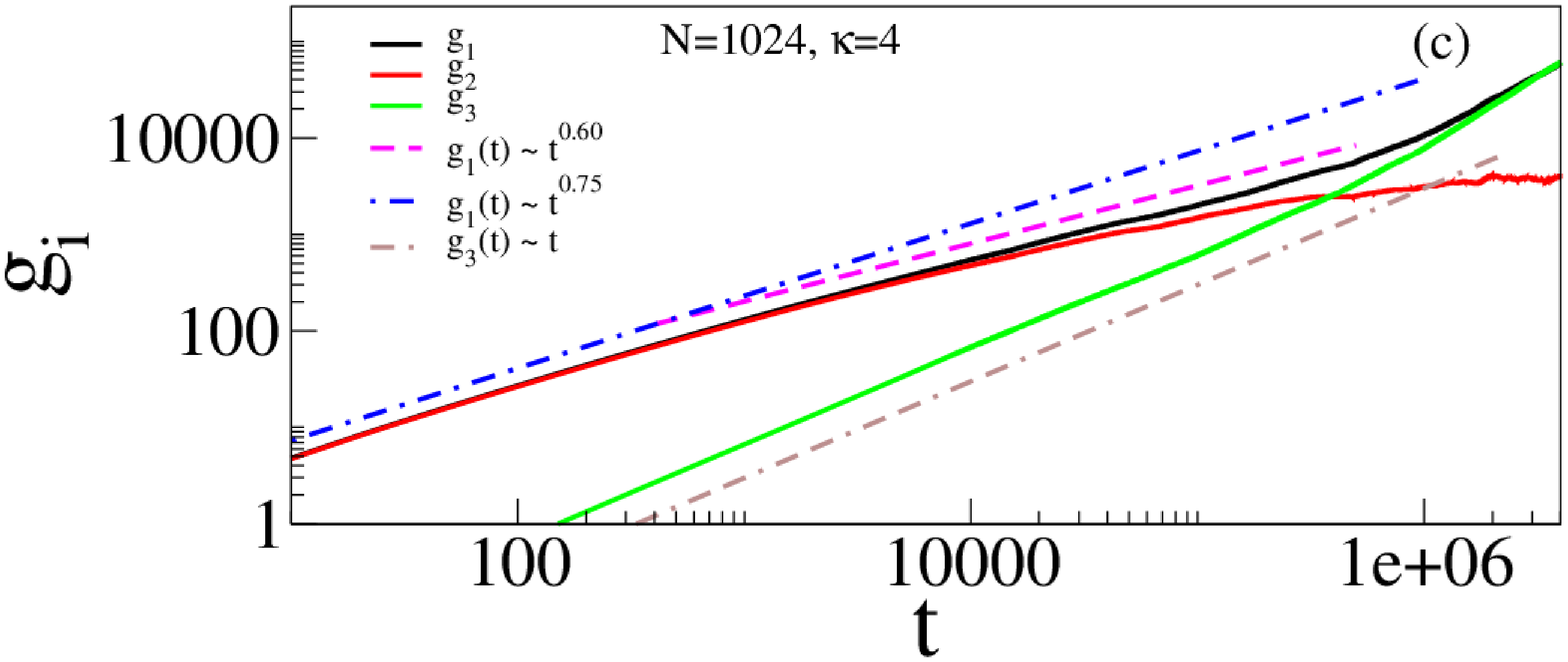}\\
\caption{\small (a) Plot for $g_1(t)$ (black), $g_2(t)$ (red) and $g_3(t)$ (green) as a function of time on a log-log scale for 
chain length $N=512$ and $\kappa = 2.0$.
The blue and magenta dashed lines correspond to straight lines $g_1(t) = At^{0.75}$,  and $g_1(t) = Bt^{0.60}$, respectively, 
where $A$ and $B$ are constants. (b) same but for $N=1024$ and $\kappa = 2.0$. (c) same but for $N=1024$ and $\kappa = 4.0$.
Note that for a fully flexible chain the slope of 
the curve $\mathrm{log}(g_i)\;\mathrm{versus}\;\mathrm{log}(t)$ would monotonously increase with time, unlike the present case.} 
\label{kappa_2_4}
\end{figure}

\begin{figure}[htp!]                
\centering
\includegraphics[width=0.89\columnwidth]{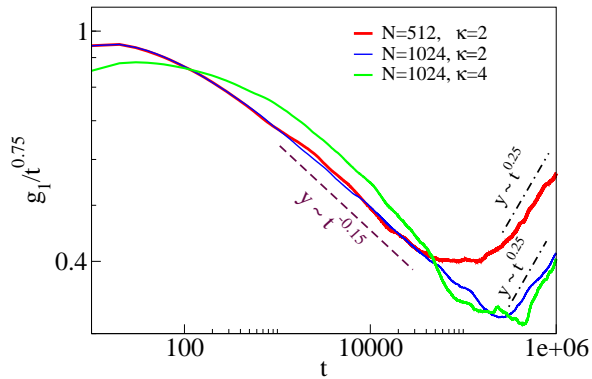}\\
\caption{Log-log plot of $g_1(t)/t^{0.75}$ as a function of $t$ corresponding to the plots of Fig.~\ref{kappa_2_4}(a) (red), (b) (blue) and (c) (green)
respectively. In each graph the minimum occurs in the intermediate regime characterized by $t^{0.6}$. } 
\label{g1_min}
\end{figure}

 
We now show data for the double crossover to support our scaling analysis for $\kappa \ne 0$. In particular we show data for chain length $N=512$ with $\kappa = $2.0 and $N=1024$ with $\kappa = $2.0, 4.0 respectively. Unlike Fig.~\ref{kappa0} for $\kappa=0$, 
plots for $g_1(t)$, $g_2(t)$ shown in Fig.~\ref{kappa_2_4} are characterized by a $t^{0.75}$ slope which then crosses over to the regime characterized by 
$g_1(t) \sim g_2(t) \sim t^{0.6}$ and that $g_1(t)$ eventually merges with $g_3(t)$. 
When we compare Fig.~\ref{kappa_2_4}(a) and  Fig.~\ref{kappa_2_4}(b) consistent with the prediction of scaling 
theory we observe that the extent of the $t^{0.75}$ region in both the graphs are the same as they have the same value of $\kappa=2$, although the chain 
lengths are different. Likewise, comparing Fig.~\ref{kappa_2_4}(b) and  Fig.~\ref{kappa_2_4}(c) we note that since both plots have the same chain length the beginning 
of the second crossover occur almost at the same time ($\tau_2 \propto N^{2.5}$) but since the latter chain is twice 
as stiff it has a wider $t^{0.75}$ regime resulting in a narrower span of $t^{0.6}$ regime.  
Thus Fig.~\ref{kappa_2_4} unambiguously confirms predictions from the scaling theory. We observe that at early time  
$g_1(t) = g_2(t)$; however the width of the region $g_2(t) \sim t^{0.6}$ is narrower than that of $g_1(t) \sim t^{0.6}$ 
and it exhibits a slightly lower value of slope before saturation. 
Considering that the crossovers are broad, we believe that this is due to finite size effect as it is evident if we compare Fig.~\ref{kappa_2_4}(a) 
and Fig.~\ref{kappa_2_4}(b). In the latter case, which is for a larger chain length, the difference between $g_1(t)$ and $g_2(t)$ are much smaller and we notice that the difference begins at a later time. Ideally for very large system the point when $g_1(t)$ would change 
its slope towards $g_1(t) \sim t$, $g_2(t)$ also tend to saturate. The sharpness of this feature would require a much larger system. 
We also note the same feature in Fig.~\ref{kappa0}. 
But the plots for $g_1(t)$ quite clearly show three distinct scaling regimes of $g_1(t) \sim t^{0.75}$ crossing over to $g_1(t) \sim t^{0.6}$ and then merging 
with $g_3(t) \sim t$ at late times. We have further confirmed the existence of this double crossover by plotting 
 $g_1(t)/t^{0.75} \sim t$ as shown in Fig.~\ref{g1_min}. The existence of an initial plateau ($ t < \tau_1$), followed by a decay ( $\propto t^{-0.15}$), 
and of a minimum (near $\tau_2$) before the diffusion ($\propto t^{0.25}$) starts further 
demonstrates quite conclusively that the exponent changes from $t^{0.75} \rightarrow t^{0.6} \rightarrow t$.
We would like to mention that because of the width of the $t^{0.6}$ regime becomes narrower for stiffer chains we were unable to see this regime 
unambiguously in simulation of shorter chains and/or larger $\kappa$ ({\em e.g.}, for $N=512$ and $\kappa \ge 4$ ). 
While for $N=512$ the double crossover is clear for 
$\kappa=2.0$, but becomes ambiguous for $\kappa \ge 4.0$ which required an increased chain length of $N=1024$. As expected, the crossovers 
are rather gradual, spread out over a decade in time $t$ 
each, and hence for chains that are not long enough the existence of these regimes is missed in previous work.\par
\begin{figure}[ht!]                
\centering
\includegraphics[width=\columnwidth]{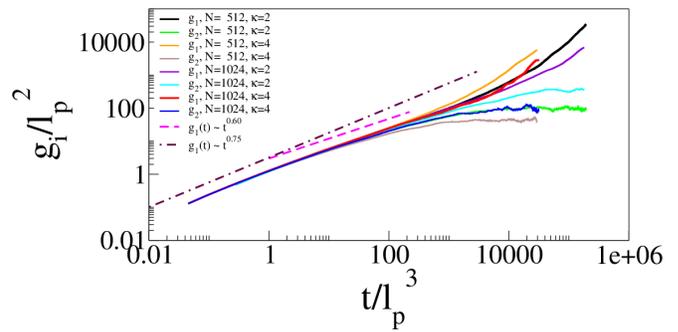}
\caption{\small Plot for $g_1(t)/\ell_p^2$ and $g_2(t)/\ell_p^2$ as a function of $t/\ell_p^3$ on a log-log scale for chain lengths $N = 512$, 1024 with $\kappa=2.0$, 4.0 respectively. The dot-dashed and dashed lines correspond to slopes 0.75 (maroon) and 0.6 (magenta) respectively.}
\label{g1g2}
\end{figure}

The interplay of Rouse modes and the bending modes with respect to the monomer dynamics of semiflexible chains was considered in early work by Harnau {\em et al.}~\cite{Harnau1}, in the framework of a Rouse model generalized by 
higher order terms to account for chain stiffness. They ignored excluded volume, and considered a single chain length, attempting to model $C_{100}$ 
alkanes in a melt. They found that their results were neither consistent with the Rouse behavior ($g_1(t) \sim t^{1/2}$) nor with the power law 
due to bending modes ($g_1(t) \sim t^{3/4}$ ). In our view, the chain length studied in this work was too 
small to observe both power laws separately, rather all their data fall in a regime of smooth crossover.
Having established the double-crossover we now further investigate the consequence of scaling prediction that the first crossover occurs 
at time $\tau_1 \propto \ell_p^3$ when $g_1(\tau_1) \propto \ell_p^2$. Fig.~\ref{g1g2} shows a plot of $ g_1(t)/\ell_p^2$ as a function of rescaled time $t/\ell_p^3 $ which shows data collapse 
for $g_1(t)/\ell_p^2 \le 1.0$ and $t/\ell_p^3 \le 1.0$ for various combinations of chain length $N$ and $\kappa$ confirming   
the length and the time scales for these crossovers. Note that the scaling theory of section~\ref{scaling_section} implies 
\begin{equation}
g_i(t)/\ell_p^2=\tilde{g}(t/\tau_1, t/\tau_2),
\label{gtilde}
\end{equation}
where $\tau_1$, $\tau_2$ are the times defined in Eqn.~\ref{tau1} and \ref{tau2}. 
If the first argument of the scaling function $\tilde{g}$, $t/\tau_1$ is small in comparison to unity, 
we can approximate Eqn.~\ref{gtilde} as $g_1(t)/\ell_p^2 \approx \tilde{g}(t/\tau_1, 0) \propto \left( t/\tau_1\right)^{3/4}$, 
which reduces to Eqn.~\ref{Granek}. For $t/\tau_1>1$ we can rewrite the scaling function as 
\begin{equation}
g_i(t)/\ell_p^2=\tilde{\tilde{g}}(t/\tau_1, \tau_2/\tau_1)=\tilde{\tilde{g}}(t/\tau_1, \left(L/\ell_p\right)^{5/2}).
\label{g_2tilde}
\end{equation}
Note that $\tau_2/\tau_1$ remains constant when we increase  $N$ and $\ell_p$ by the same factor: this
observation explains that the scaling of the data in Fig.~\ref{g1g2} encompasses the full range of times. \par
\section{Summary and Discussion}
In conclusion, we have studied conformations, fluctuations, and crossover dynamics of a swollen semiflexible chain in 2d. We first developed a scaling 
theory which generalizes early time monomer dynamics of a fully flexible chain for a semiflexible chain characterized by its contour 
length and the persistence length.  We predict a double crossover which arises due to the presence of an additional length scale introduced through 
the chain persistence length. Monomer dynamics up to a length scale $\ell_p$ is independent of chain length and is characterized by the $t^{0.75}$ power law.
At a later time $ t > \tau_1$ when the size of the fluctuations becomes bigger than $\ell_p$ the dynamics begin to look like that of a fully flexible 
chain and characterized by the well known $t^{2\nu/1+2\nu}$ growth. Both of these exponents have been discussed in the literature separately 
but have not been emphasized that before the entire chain reaches purely diffusive regime, there ought to be two and not one crossover, the first 
crossover differentiates chains of different stiffness. Previously the dynamics of monomer MSD of 
semiflexible polymers has also been studied by Harnau {\em et al.}~\cite{Harnau1}, using a Rouse-type model generalized to include chain stiffness. They saw a gradual crossover, in between bending modes and Rouse modes, but did not consider the scaling description of the crossover. Note that excluded volume effects were absent in their model, and hence it is not applicable in $d=2$ dimensions. \par
Motivated by recent lattice MC results for a swollen chain in 2d predicting the absence of a Gaussian regime we undertook similar studies in 2d continuum 
using BD simulation. While checking our data for the RMS end-to-end distance for chain of different contour length and persistence length we discovered 
that we regain the well known results for the end-to-end distance due to Schaefer, Joanny, and Pincus~\cite{Pincus_MM_1980} and Nakanishi~\cite{Nakanishi_1987} provided we use the definition 
of the persistence length given by either the lattice or continuum version of the Kratky-Porod WLC model. We explain this by noting that the persistence 
length being a local property of a chain does not depend on the EV interaction. This is further reassured when we note that the persistence length calculated 
this way does not depend on the chain length unlike well used textbook definition of persistence length where the projection of the end-to-end vector 
on the first bond is used as the definition and does depend on the chain length. Therefore, we emphasize that the latter definition needs to be discarded.\par 
We also confirm the absence of the Gaussian regime in the continuum bead-spring model where the swollen chain for $L/\ell_p \ll 1$ behaves like a rod
and thereafter always behaves like a swollen chain. Considering that there are increased number of activities to explore the properties of biomolecules 
on a surface, our result (Fig.~\ref{rn}) will be extremely valuable to analyze the experimental data correctly for stiff molecules on flat surfaces.
It has not escaped our attention that many such reported analyses are still done using the WLC model and/or calculating the chain persistence length from 
projected end-to-end to the first bond.\par 
Transverse fluctuations in a stiff chain has been addressed analytically in the literature only in the extremely stiff chain limit where one finds that it is described by 
the roughening exponent. Analytic calculations for moderately stiff chains are hard to carry out, and to date there are no results for transverse fluctuations
spanning the entire regime from a stiff to a fully flexible chain. We have numerically obtained this result and pointed out that 
the appropriate length variable to analyze the data
is to use the persistence length as the unit of length. When we use $L/\ell_p$ as the length scale to plot transverse fluctuation 
we discover the non-monotonic 
behavior of this fluctuation reaching a maximum for some $L \gtrapprox \ell_p$. We point out that this universal scaling of the 
transverse fluctuation can be used to measure the persistence length of the chain. Another accompanying consequence of the absence of 
Gaussian regime for a swollen chain in 2d is the decay of the bond correlation function which exhibits a power law decay. 
Again, by choosing the normalized contour segment $s/\ell_p$ as the appropriate variable we regain the exponent $\beta = 0.5$ which describes the decay 
of bond autocorrelation for a fully flexible chain. We must point out that many of these results and analyses on chain conformations 
and equilibrium fluctuations point to a common theme. In the limit $L/\ell_p \gg 1$ we recover the expected behavior of a fully flexible 
chain and chains with different stiffness exhibit universal scaling behavior when persistence length is 
chosen as the unit of length.  \par
This general idea extends to our study of monomer dynamics as well. 
We have provided a new scaling theory of monomer dynamics for semiflexible polymers in 2d. Our theory predicts novel crossover dynamics at an intermediate 
time when the fluctuations of the monomers $g_1(\tau_1) \sim \ell_p^2$. Around this time the monomer dynamics become the same as that of a fully flexible 
swollen chain characterized by $g_1(t) \sim t^{2\nu/(1+2\nu)} =  t^{0.6}$ in 2d . The theory expands the existing scaling 
theory for monomer dynamics for a WLC and that of a fully flexible chain to include the effect of the chain persistence length. 
Fully flexible swollen chains are self-similar objects, while a polymer segment up to its own 
persistence length is not. Therefore, it is expected that for length scales up to $\ell_p$ the dynamics will have different characteristics due to bending modes arising 
out of the chain stiffness. The EV effect is almost negligible for the $t^{0.75}$ regime and therefore, our result is the same as that of previous studies using 
WLC Hamiltonian~\cite{Granek_1997,Maggs_MM_1993}. For the $t^{0.6}$ regime originating from the EV effect, 
where the monomer dynamics are governed by Rouse relaxation of a fully flexible chain, 
our theory elucidates the exact role of chain persistence length neither contained in WLC model nor studied before. 
We also validate our new scaling theory by extensive BD simulation results.\par
A subtle issue concerns the limit $\kappa$ towards infinity while keeping the contour length $L$ fixed. Then transverse motions of the monomers relative to each other, in a coordinate system where the $x$-axis is fixed along the rod-like polymer, are completely suppressed. In the ``laboratory coordinate system", however, the rod still can make random transverse motions, namely rigid 
body rotations and translations. However, in addition to those motions still
motions of the monomers relative to each other along the axis of the rod are possible. These motions may give rise to a transient $t^{1/2}$ behavior, as a model 
calculation for a one-dimensional harmonic chain shows~\cite{huang_1d}.
However, our data for $g_i(t)$ for large $\kappa$ due to the smoothness of crossovers did not allow to clearly separate this mode of motions from the displacements due to transverse fluctuations.\par
In the present manuscript we have ignored HD interactions as they are not significant for 2d swollen chains on a substrate.
However we now present simple estimates of generalization of our results in 3d and/or in presence of HD interactions which 
will be relevant for a 3d swollen chain. 
In the free draining limit, the results $t^{0.75}$  
will remain the same in 3d~\cite{Granek_1997,Maggs_MM_1993}, but the intermediate Rouse relaxation regime 
will be characterized by 
$t^{2\nu/(1+2\nu)} = t^{0.54}$ ($\nu = 0.59$ in 3d), for the case where the EV is relevant 
({\em i.e.}, MSDs exceeding $R^{\ast 2}=\ell_p^2/b^2$). For MSD in between $\ell_p^2$ and $R^{\ast 2}$ Gaussian 
behavior prevails, $\nu=1/2$, and hence $g_1(t)\propto t^{1/2}$ in that regime. The crossover between flexible and 
stiff chain dynamics in 3d in the free draining limit was studied by Steinhauser {\em et al.}~\cite{Steinhauser}, 
but no scaling analysis is done. \par
Replacing Rouse relaxation by Zimm relaxation one immediately 
sees that in presence of HD interaction the intermediate regime is characterized by $t^{2\nu/3\nu}=t^{2/3}$~\cite{Hinczewski,Netz_EPL_2009,Hinczewski_PhysicaA}. 
Notice that in this case $\nu$ cancels out and this relaxation should be the same in 2d and 3d.  
However, as shown by Hinczewski and Netz ~\cite{Netz_EPL_2009,Hinczewski_PhysicaA}, very complicated crossovers occur in this case.\par
We now make some comments about some recent experiments to study monomer dynamics. This is typically done using FCS where a tagged monomer 
can be directly watched in real time. However, as has been mentioned by Petrov {\em et al.}~\cite{Petrov} that since the $t^{0.75}$ regime 
or the intermediate regimes (either $t^{0.5}$ or $t^{2/3}$) occur at much shorter time scales compared to the longest relaxation time, 
unless extreme caution is taken for the measurement of MSD of a labeled particle, the interpretation can be misleading, especially for shorter 
DNA fragments~\cite{Shusterman_PRL_2004}. For dsDNA of length $10^2 - 2\times 10^4$ base pairs (which is equivalent to $L/\ell_p \sim 0.7-140)$ fluorescence correlation spectroscopy (FCS) studies of Petrov {\em et al.} observed the Zimm regime characterized by a $t^{2/3}$ power law. However, a clear demonstration of the double crossover is lacking 
in recent experiments with biopolymers~\cite{Sackmann_NJP,Goff_PRL_2002,Capsi,Shusterman_PRL_2004,Petrov}. This is 
partly due to lack of resolution of the experiments and partly due to the fact that   
lacking any theoretical predictions for this phenomenon, researchers did not specifically investigate the precise behavior 
of MSD before the onset of overall chain diffusion very carefully. We will provide some physical arguments why the experimental detection 
can be hard: a simple calculation for Fig.~\ref{phase} shows that in order for the width of the $t^{0.75}$ and $t^{0.60}$ to 
be equal (in logarithmic scale) one needs $N = \ell_p^{2.2}$ in 2d. In other words for a stiffer chain one needs 
a very long chain to see the $t^{0.60}$ regime. Indeed in our simulation we found (not shown here) that for $\kappa = 16$, 32, and 64, 
the results with chain length up to $N=512$ are largely dominated by the $t^{0.75}$ regime and we did 
not clearly see the  $t^{0.60}$ regime. It is only after we lowered the value of $\kappa$ and used longer 
chain ($N=1024$), we identified these two regimes quite conclusively (Fig.~\ref{kappa_2_4}).
We suspect that the same might happen in experiments~\cite{Sackmann_NJP}. For extremely stiff chains 
the $t^{0.6}$ (or $t^{0.54}$ in 3D) region can be extremely narrow and could either be missed or the rather smooth double 
crossover might be mistakenly interpreted as a single crossover (with $t^{2/3}$ in 2d). 
Therefore, we believe that these results will not only promote new experiments but will be extremely valuable 
in identifying and interpreting different scaling regimes for the monomer dynamics of semiflexible polymers.\par
\newpage
\begin{acknowledgments}
The research has been partially supported (AH) by the UCF Office of Research \& Commercialization and 
the UCF College of Science SEED grant. AB acknowledges travel support from Deutsche Forschungsgemeinschaft at the 
Institut f\"ur Physik, Johannes Gutenberg-Universit\"at, Mainz. We thank both the referees for their constructive 
criticism and comments.
\end{acknowledgments}

\end{document}